\documentclass[twoside,twocolumn]{article}

\usepackage[sc]{mathpazo} % Use the Palatino font
\usepackage[T1]{fontenc} % Use 8-bit encoding that has 256 glyphs
\usepackage{lmodern} % SH added this
\usepackage{microtype} % Slightly tweak font spacing for aesthetics
\usepackage{amssymb} % ES added for rightleftarrows

\usepackage[english]{babel} % Language hyphenation and typographical rules

\usepackage{graphicx}
\usepackage{color}
\usepackage[top=2cm,bottom=2cm,left=1.5cm,right=1.5cm,columnsep=20pt]{geometry} % Document margins
    {\large\scshape Computational Details%
    \par\medskip\normalfont\normalsize}%
    {}%
\newenvironment{acknowledgment}%       New acknowledgment environment
    {\large\scshape Acknowledgment%
    \par\medskip\normalfont\normalsize}%
    {}%
    {\large\scshape Supporting Information%
    \par\medskip\normalfont\normalsize}%
    {}%
\usepackage{multirow} % SH added for table
\usepackage[normal, small,labelfont=bf,up,textfont=it,up]{caption} % SH added this (180704)
\usepackage{booktabs} % Horizontal rules in tables

\usepackage{lettrine} % The lettrine is the first enlarged letter at the beginning of the text

\usepackage{enumitem} % Customized lists
\setlist[itemize]{noitemsep} % Make itemize lists more compact

\usepackage{abstract} % Allows abstract customization
\usepackage{titlesec} % Allows customization of titles
\renewcommand\thesection{\Roman{section}} % Roman numerals for the sections
\renewcommand\thesubsection{\roman{subsection}} % roman numerals for subsections
\titleformat{\section}[block]{\large\scshape\centering}{\thesection.}{1em}{} % Change the look of the section titles
\titleformat{\subsection}[block]{\large}{\thesubsection.}{1em}{} % Change the look of the section titles

\usepackage{fancyhdr} % Headers and footers
\usepackage{titling} % Customizing the title section

\usepackage{hyperref} % For hyperlinks in the PDF

\pretitle{\begin{center}\Large\bfseries} % Article title formatting
\posttitle{\end{center}} % Article title closing formatting

\def\bea{\begin{eqnarray}}
\def\eea{\end{eqnarray}}
\def\ben{\begin{equation}}
\def\een{\end{equation}}
\def\benu{\begin{enumerate}}
\def\enu{\end{enumerate}}

\def\bei{\begin{itemize}}
\def\eei{\end{itemize}}
\def\beit{\begin{itemize}}
\def\eit{\end{itemize}}
\def\benu{\begin{enumerate}}
\def\enu{\end{enumerate}}

\def\n{n}

\def\sss{\scriptscriptstyle\rm}

\def\1var{(\bx_1...\bx\N)}

\def\br{{\bf r}}

\def\bx{{x}}

\def\x{_{\sss X}}

\def\xc{_{\sss XC}}

\def\N{_{\sss N}}

\def\HF{^{\rm HF}}

\def\LDA{^{\rm LDA}}

\def\PBE{^{\rm PBE}}
\def\DFA{^{\rm DFA}}

\def\sph_int{ {\int d^3 r}}

\usepackage[table]{xcolor}
\usepackage{amsmath}
\usepackage{graphicx}

\usepackage[]{quoting}
\usepackage{multirow} % SH added for table
\usepackage[table]{xcolor}
\def\d{_{\sss D}}
\def\f{_{\sss F}}

\usepackage{xr}
\usepackage{float}

\newcounter{change}
\title{Improving results by improving densities: Density-corrected density functional theory}

\author{%
\textsc{Eunji Sim$^{a,}$\thanks{esim@yonsei.ac.kr}, Suhwan Song$^a$, Stefan Vuckovic$^{b,c}$, and Kieron Burke$^d$} \\ % name
\normalsize $^a$Department of Chemistry, Yonsei University, 50 Yonsei-ro Seodaemun-gu, Seoul 03722, Korea \\ % institution
\normalsize $^b$Institute for Microelectronics and Microsystems (CNR-IMM), Via Monteroni,Campus Unisalento, 73100 Lecce, Italy \\
\normalsize $^c$Department of Chemistry\&Pharmaceutical Sciences and Amsterdam Institute of Molecular and Life Sciences (AIMMS), \\
\normalsize Faculty of Science, Vrije Universiteit, De Boelelaan 1083, 1081HV Amsterdam, The Netherlands\\
\normalsize $^d$Departments of Chemistry and of Physics, University of California, Irvine, CA 92697, USA \\  % institution
}
\date{\today} % Leave empty to omit a date

\renewcommand{\maketitlehookd}{%
\begin{abstract}
\noindent 
DFT calculations have become widespread in both chemistry and materials, because
they usually provide useful accuracy at much lower computational cost than wavefunction-based
methods.   All practical DFT calculations require an approximation to the unknown
exchange-correlation energy, which is then used self-consistently in the Kohn-Sham scheme
to produce an approximate energy from an approximate density.   
Density-corrected DFT is simply the study of the relative contributions to the total energy error.
In the vast majority of DFT calculations, the error due to the approximate density is negligible.
But with certain classes of functionals applied to certain classes of problems, the density
error is sufficiently large as to contribute to the energy noticeably, and its removal leads to
much better results.   These problems include 
reaction barriers, torsional barriers involving
$\pi$-conjugation, halogen bonds, radicals and anions, 
most stretched bonds,
etc. In all such cases, use of a more accurate density significantly improves performance, and
often the simple expedient of using the Hartree-Fock density is enough.   This article explains
what DC-DFT is, where it is likely to improve results, and how DC-DFT can produce more accurate
functionals.  We also outline challenges and prospects for the field.
\end{abstract}
}

\begin{document}
\sf
\maketitle

\section{Introduction}
\label{sec:intro}

Density functional calculations have become ubiquitous in modern chemistry 
and materials science since the award of the 1998 Nobel prize in chemistry.\cite{K99}
There are now many computer codes available for performing such 
calculations.\cite{T15,G16,qchem,VASP,PYSCF,ORCA}
It is a straightforward matter to choose a basis set and an approximate
functional, and calculate an interesting property, such as a reaction
barrier, bond length, or dipole moment.    But it requires judgment
and experience to choose wisely.\cite{B12}  Ensuring the quantity is converged with
respect to basis is relatively simple.  Given hundreds of possible DFT
approximations available in a code, the choice can be difficult.\cite{RCFB08}

There are myriad approaches to constructing exchange-correlation (XC) approximations, varying from
appeals to general principles of quantum mechanics to fits 
to large databases.\cite{PGB15,MH17,GHBE17,VT20} 
Modern approximations include generalized gradient approximations (GGA), hybrids,
range-separated functionals, the random phase approximation and variants thereof,
dispersion corrections of at least three distinct flavors, double-hybrids, 
and many, many more.\cite{MH17,MS19,J21,Ja17}
All over the world, theorists of many different backgrounds work at improving (or at least,
expanding) on our current choices, either with improved accuracy, lower computational cost,
or greater reliability.\cite{NABM19}

In each of the countless DFT calculations performed worldwide each year,
the Kohn-Sham (KS) equations\cite{KS65} are iterated to
a self-consistent (SC) electronic density and orbitals, and the total energy
of the system is reconstructed with these final quantities.   By definition,
this process finds the unique\cite{HK64} density that minimizes the approximate
energy.   All components of that energy are exactly determined, apart from the
notorious XC energy.   It is that piece which is approximated
in DFT and whose derivative appears in the KS equations as the XC potential.

Thus, whatever choice of XC is made, it is actually used twice in the calculation.  Once in finding
the density and a second time in finding the energy, so that neither is quite correct.  
As the foundation of DFT is to consider the energy as a functional of the density\cite{HK64}, we may
write the error in any self-consistent KS calculation as:
\ben
\Delta E = \tilde E[\tilde \n]-E[n]
\een
where $n(\br)$ is the exact density and $E[n]$ is the exact functional, while tildes denote
approximate quantities.  In most practical calculations,
modern XC approximations yield excellent approximate densities\cite{KSB13}, so that the
energy error would barely
change if the approximation were evaluated on the exact density.

\begin{figure*}[htb]
\centering
%\vskip 5cm
\includegraphics[width=1.95\columnwidth]{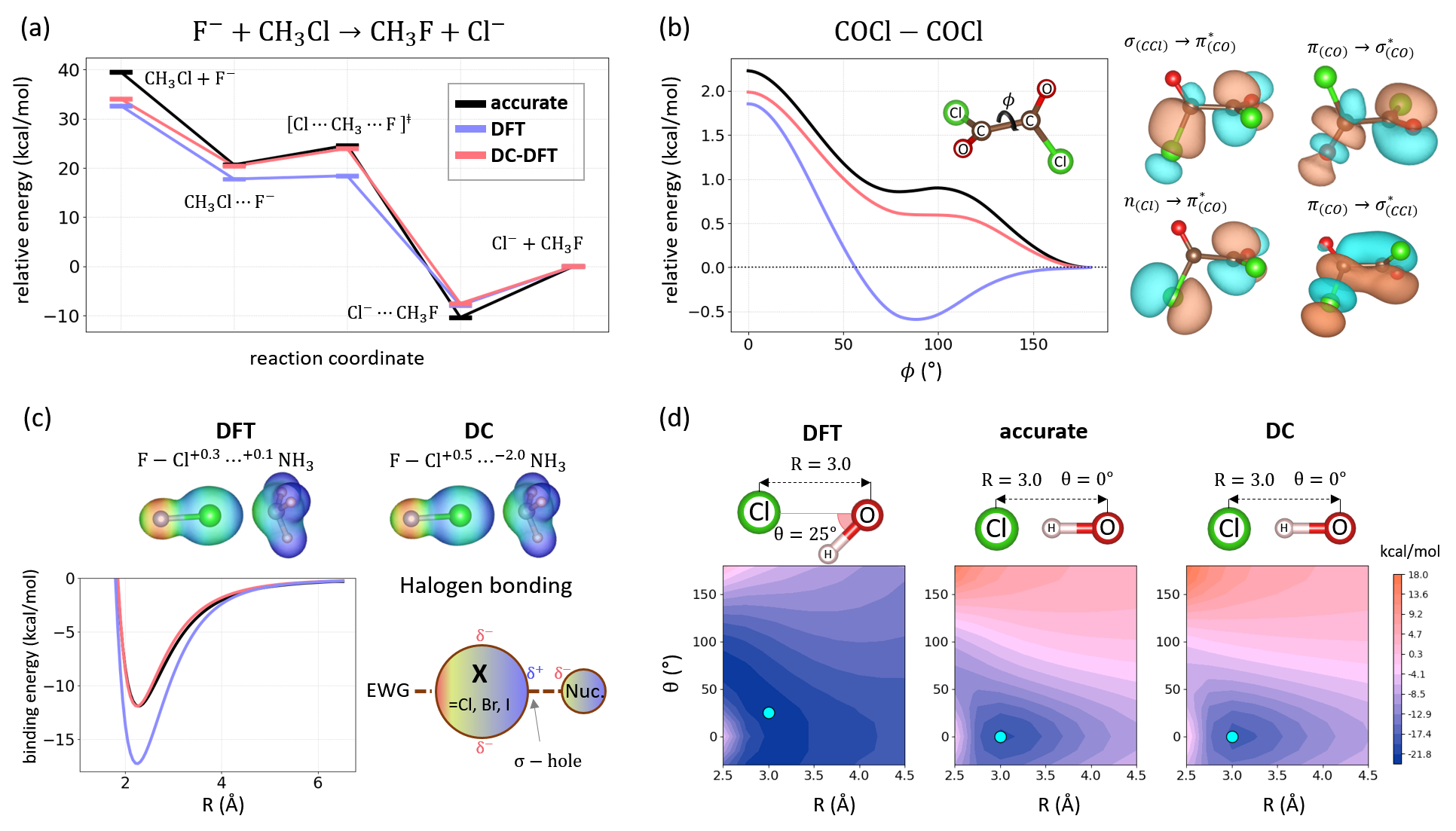}
\caption{
Representative cases 
where standard self-consistent DFT fails, but DC-DFT restores the correct
energetic and/or geometric information.
(a) Reaction coordinate diagram of a textbook S$_N$2 reaction, interconversion between negatively charged F and Cl ions.
(b) Torsional rotation energy profile and 
selected NBO donor--acceptor pairs of oxalyl chloride.\cite{NCSB21}
(c) Intermolecular interaction between NH$_3$ and ClF interacting via halogen bonds; 
Electrostatic attraction between a partially negatively charged nucleophile (Nuc.)
and a partially positively charged halogen (X) bonded to an electron withdrawing group (EWG).\cite{KSSB18}
(d) The potential energy surface of HO$\cdot$Cl$^-$
as a function of the O--Cl$^-$ distance and
the H--O--Cl$^-$ angle.\cite{KSB14}
Here, DC-DFT represents the calculation result using the DC(HF)-DFT method.
A detailed description of DC(HF)-DFT is provided later.
}
\label{fgr:super}
\end{figure*}

It is certainly extremely convenient to use the self-consistent solution density.  It
is easily computed from the KS equations.  By being self-consistent, many important properties,
especially those depending on derivatives of the energy, are much simpler and many additional
terms need not be calculated.   This is so convenient that essentially all modern codes use
the self-consistent density in almost all circumstances.   However, this was not always so.
In the earliest days, 
the Hartree-Fock (HF) density was often used instead.\cite{LC74,CS75,GJPF92a,GJPF92b,JGP92}
Mostly, it was used
as a matter of convenience, so as to avoid needing to do a self-consistent calculation, in the
belief that it mattered little.  Later, the HF density was used as a matter of principle, to
compare functionals against each other without having to worry about changes 
in the density.\cite{GJPF92a,GJPF92b,JGP92}
It was even presciently noted that, in some cases, it really did seem to matter, and in those
cases, it was often better to use HF densities.\cite{CN93,OB94,JS08}

This article shows that, in fact, it really does matter, both theoretically and very practically.
Until about 10 years ago, no careful, systematic analysis had been performed on this question.
In fact, every single KS-DFT calculation ever run can be analyzed, to separate its functional
error (energy error made on exact density) from its density-driven error (the remainder).  
Surprisingly large classes of calculation, such as typical reaction barriers, contain 
significant density-driven errors with standard functionals, such as B3LYP.
One of the major reasons for this is the 
over-delocalization of charges and spins due to 
semilocal XC approximation.\cite{CMY08,LZSY17}
These errors are typically substantially reduced
by using the HF density instead of the self-consistent density.
Even highly
accurate (and expensive) DFT approximations such as double-hybrids can be improved by
separating out these two error sources in their design.

Figure~\ref{fgr:super} is a panoply of calculations where the density really matters.
In every case, when self-consistent densities are replaced by HF densities,
the energy errors drop by a substantial margin.  In panel (a),
we show an energy diagram for a textbook S$_N$2 reaction.  
Starting from either reactants or products, negatively charged complexes are formed barrierlessly, 
while the interconversion between the two involves a barrier of
$\sim$35~kcal/mol in the backward and $\sim$5~kcal/mol in the forward directions.  
Standard DFT provides
reasonable reaction energies, but fails badly for barrier heights
in both directions for the complex interconversion.
DFT underestimates the backward barrier height by about 10~kcal/mol, implying
a reaction many orders of magnitude faster than reality.
The barrier height is indeed smaller in the forward %other 
direction, but standard DFT yields no barrier at all.
A long time ago it has been demonstrated that the use of HF densities 
fixes failures of DFT for barrier heights, and so does here.\cite{JS08,VPB12}
For the backward barrier, density-corrected DFT (DC-DFT) reduces the error of DFT by about 6~kcal/mol, whereas 
the DC-DFT forward barrier height matches the reference.
%\vs{[A small thing: Because the reactant is here on the right (where energy is zero), 'forward' and 'backward'  might be confusing.  Maybe better as was before: from left to right, and from right to left. or put the reactant on the left.]}
%\es{[Chemists define forward/backward as the way chemical equations are written, so forward/backward is less confusing. People usually set 0 on the reactants, but we did set the product to 0 in order to make DC-DFT look better in this system.]}

Panel (b) demonstrates the power of using DC-DFT to fix the failures of DFT for
difficult torsional barrier heights, whose accurate predictions play a
crucial role in describing a range of chemical processes (e.g., selectivity, protein folding, molecular electronics, etc.).\cite{NCSB21}
Most torsional barriers are very accurate with standard DFT (errors below 1~kJ/mol), but
barriers of a single bond participating in $\pi$-conjugation are particularly problematic for DFT.\cite{GHBE17,KCK97}
For the oxalyl chloride shown, the standard DFT energy diagram is qualitatively wrong,
incorrectly predicting that the {\em perpendicular} conformation is more stable ($\phi \sim 90^\circ$)
than the {\em trans} conformation ($\phi=180^\circ$), where $\phi$ is the torsional angle depicted in panel (b).  
DFT also finds that there is no barrier upon conversion from {\em trans} to {\em perpendicular}.  
Using DC-DFT with HF densities, these barriers become far more accurate as shown.

For some weak interactions, such as halogen bonds, DC-DFT greatly improves over
its self-consistent counterpart.  
The binding energy curve for one halogen bonded complex is shown in panel (c).\cite{KSSB18} 
Standard DFT overbinds the complex by about 50~$\%$ at equilibrium, whereas 
the DC-DFT binding curve is almost indistinguishable from the reference.  
In contrast to DC-DFT, 
dispersion corrections (such as the commonly used D3\cite{GAEK10}) cannot fix bad
DFT densities, and their addition has almost no effect on the DFT binding curve in panel (c). 
Despite these large improvements in energetics when HF densities are used in place of self-consistent
densities, their electrostatic potentials are almost identical, as shown.
From the DC-DFT perspective, there is no need to stare at density or electrostatic potential plots
to decide which density is better.
DC-DFT measures the accuracy of densities directly in terms of their impact on the energy, the
quantity that really matters.  Even the tiny differences visible in the electrostatic potentials
can be measured.

Finally, panel (d) reminds us of one of the
earliest successes of DC-DFT -- the description of odd-electron radical complexes,
which play important roles in atmospheric and environmental chemistry, cell biology, 
etc.\cite{KSB14}
In panel (d), we compare 
potential energy surface 
for the HO$\cdot$Cl$^-$ complex by varying the $R$ distance and $\theta$ angle, as shown.
Self-consistent DFT fails badly in simulating the potential energy surface:
(i) it finds that the equilibrium
structure is bent instead of linear ($\theta \sim 30^\circ$ instead of $0^\circ$),
(ii) gives contour of the wrong shape leading to wrong forces, and 
(iii) gives too blue (too negative) potential energy surface.
DC-DFT again saves its self-consistent counterpart by not only yielding the correct 
linear structure of HO$\cdot$Cl$^-$ as the most stable, but also producing a far
more accurate potential energy surface and equilibrium structure.  In this way, we see that DC-DFT
not only improves DFT energetics, but also gives more accurate geometries and force fields.
For applications of the principles of DC-DFT to geometry optimization of any electronic structure method, see Ref.~\cite{VB20,V22}, which contains many surprising results about functional performance for geometry.

For the cognoscenti, in Figure~\ref{fgr:super}, all DFT calculations are with PBE, except in (a) which uses
B3LYP, and all accurate reference calculations are DLPNO-CCSD(T)-F12, except in (d), which is
simple CCSD(T).
The rest of this article is about why the basic ideas of DFT do {\em not} imply always choosing
the self-consistent density.  This is followed by a discussion of practical DC-DFT, with many
examples illustrating crucial aspects of density-sensitive systems and calculations.
We next explore some of the finer points of theory, ending with a surprise:  Although Diels-Alder
reactions are {\em not} density sensitive, functionals (double-hybrids) designed to take advantage
of DC-DFT perform better.  We end with many challenges and potential of DC-DFT.

\begin{figure}
\centering
\includegraphics[width=0.7\columnwidth]{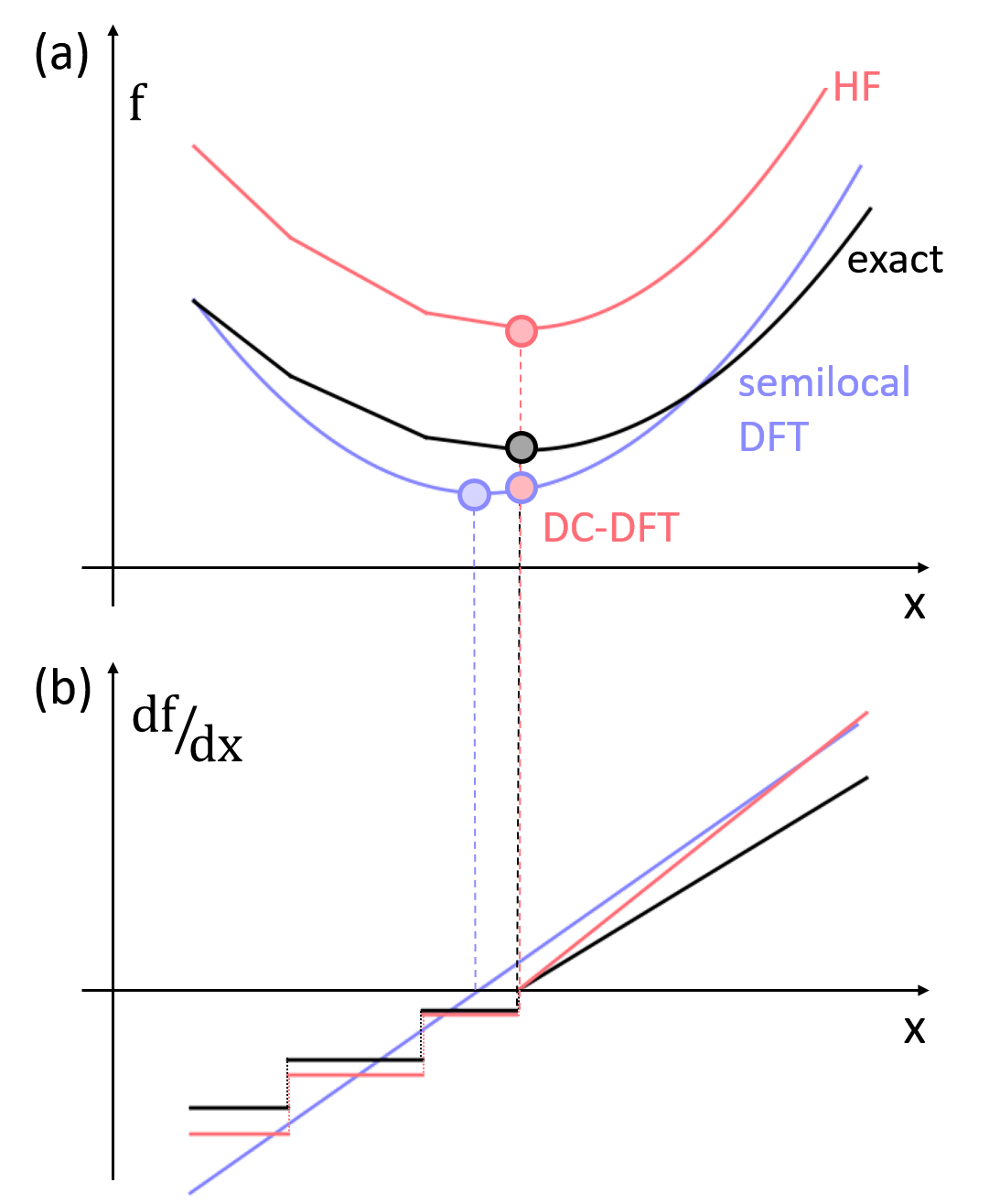}
\caption{Cartoon illustrating how semi-local DFT can be more accurate than HF everywhere, but still
produce a more accurate energy when evaluated at the HF minimum:
(a) Total energies as a function of density and (b) their corresponding derivatives.
}
\label{fgr:cartoon}
\end{figure}

%\newcommand*\mycommand[1]{\texttt{\emph{#1}}}
%\newcounter{change}
\setcounter{change}{5}

First we ask, why is this a question at all?  Surely the self-consistent density is `best' because
it minimizes the (approximate) energy functional?  It does, but because the functional is  
approximate, its minimum might well be below the true ground-state energy.  Moreover,
all useful properties are actually energy differences, and the difference between two minima 
does not obey a variational property.   One of the most well-documented failings of most
density functional approximations is that they are too smooth, especially as particle numbers
pass through integer values.\cite{Pb85,PY89}    You might object that, in reality,
all molecules have integer numbers of
electrons.   But as a bond is stretched, the exact functional develops sharp cusps that 
force integer numbers of electrons onto each fragment, while typical
DFT approximations are smooth.\cite{KPSS15}
Figure~\ref{fgr:cartoon} 
is a cartoon showing how 
function with a cusp can be well-approximated by 
a smooth one everywhere, but whose derivative is very wrong in the vicinity of that cusp.  
Semilocal XC approximations yield curves 
that are smooth everywhere, 
which causes the overdelocalization of  charges when bonds are stretched.\cite{LZCM15}
The HF energy functional depends explicitly on only occupied orbitals, 
making it often even sharper than the exact functional.  
Relative to HF, correlation includes infinite sums over orbitals, which typically dampen the cusp as the particle number changes.
%On the other hand, because of the functional's explicit dependence on orbitals, the HF
%energy functional is often even sharper than the exact functional.  Using its density
%counteracts the error made by being too smooth.

\vspace{0.1cm}\noindent{\textbf{\small{\em Basic separation into functional and density-driven errors}:}}
Having established that the self-consistent density need not yield the most accurate energy,
how then should
we decide when we might want to avoid it?   We simply decompose the energy error into two 
well-defined pieces.\cite{KSB13}
The functional error is simply the error in the energy if
we had evaluated it on the true density, not the self-consistent one.  Many would consider this
the `true' DFT error, as this is an apples-to-apples comparison.   Moreover, the beauty
of the KS scheme is that only the XC contribution to the energy is approximated.  Thus, in
any KS calculation, the functional error is entirely due to the XC approximation:
\ben
\Delta E\f = \tilde E [\n] - E[\n] = \tilde E\xc[\n]-E\xc[\n].
\een
The remainder of the energy error is called the density-driven error, 
\ben
\Delta E\d= \tilde E[\tilde\n]-\tilde E[\n],
\een
\noindent and is given by the
difference in the approximate functional on the exact and self-consistent densities.   This
is always negative for any given energy calculation.

\vspace{0.1cm}\noindent{\textbf{\small{\em Universality of energy decomposition:}}}
Thus, no matter what XC approximation you use or can afford, no matter what molecule or solid you
study, and no matter which property you extract from your KS-DFT calculation, you will have
some error, and that error is the sum $\Delta E\f + \Delta E\d$.
In the vast majority of routine calculations, the self-consistent DFT densities
are incredibly accurate, so that the density-driven term has negligible effect,
and DC-DFT will not help ($|\Delta E\d| \ll |\Delta E|$).
But, with certain classes of approximation, certain classes of
molecules, and certain properties, it has been found that the density-driven error is large  
enough to substantially contribute or even distort calculations.\cite{SSB18,VSKS19,NSSB20}
Moreover, in such cases, using a better density has led
to much better energetics.\cite{SVSB21}

With some thought, these statements would appear paradoxical.   If the functional is working
well for the system you are calculating, how could its density be wrong?   Well, this happens
because its derivative, the XC potential, is sufficiently inaccurate as to produce a sufficiently
flawed density as to mess up your energy evaluation.   
Return to Figure~\ref{fgr:cartoon} to see a good approximation
to a function whose derivative is lousy.   Doesn't a better functional automatically imply
a better XC potential?  No, it does not.  Almost all modern XC approximations have very poor-looking
XC potentials, often shifted by very large amounts relative to the exact XC potential.\cite{UG94}
Yet they still usually yield highly accurate densities in the regions where it matters.
GGA approximations to XC often have worse looking potentials than their LDA counterparts,
but nonetheless have much better energetics.\cite{LCB98,BCL98}

In the original work\cite{KSB13}, 
the term 'abnormal' was used to designate those KS-DFT
calculations whose results were contaminated by density-driven errors, and
this is a characteristic of the approximate functional, the property of interest, and
the given system.  By contaminated, one means that the error in the energy being
calculated changes substantially if the exact density is used instead.
There a small KS HOMO-LUMO gap in self-consistent DFT was identified as a signal of abnormality.
But the use of the gap as the abnormality indicator is not ideal, as some calculations
(e.g., those involving stretching of homonuclear bonds) have small HOMO-LUMO gaps without density-driven
errors.
More appropriate indicators of abnormality have been built and are detailed below.

\section{Practical DC-DFT}
\label{prac}

In practice, much of the above is just so much theorizing, as, if we need to do a DFT
calculation, we surely cannot afford to calculate the exact (or highly accurate) density.
Fortunately, we show below that, in the cases where there is a significant density-driven
error with a standard DFT calculation, very often using the HF density significantly
reduces the density-driven error.   This is presumably because HF, although yielding woefully
inadequate energetics, suffers from the reverse of the errors of most density functional
approximations.   Essentially, DFT approximations almost always include some variety of
semi-local density functional (i.e., depending on the density, its gradient, and/or its
laplacian or kinetic energy density).   These approximations tend to delocalize the density
relative to the exact one, whereas the HF density is typically overlocalized.   
This is not to say that HF densities are somehow 'better' than approximate self-consistent
DFT densities.   As discussed above, there is no well-defined meaning to being better.
All this means is exactly what is stated:  In cases where the self-consistent
approximate DFT density is unusually poor, the HF density is often more accurate in the
very precise and limited sense of yielding more accurate energetics.

\begin{figure*}[htb]
\centering
%\vskip 5cm
\includegraphics[width=1.95\columnwidth]{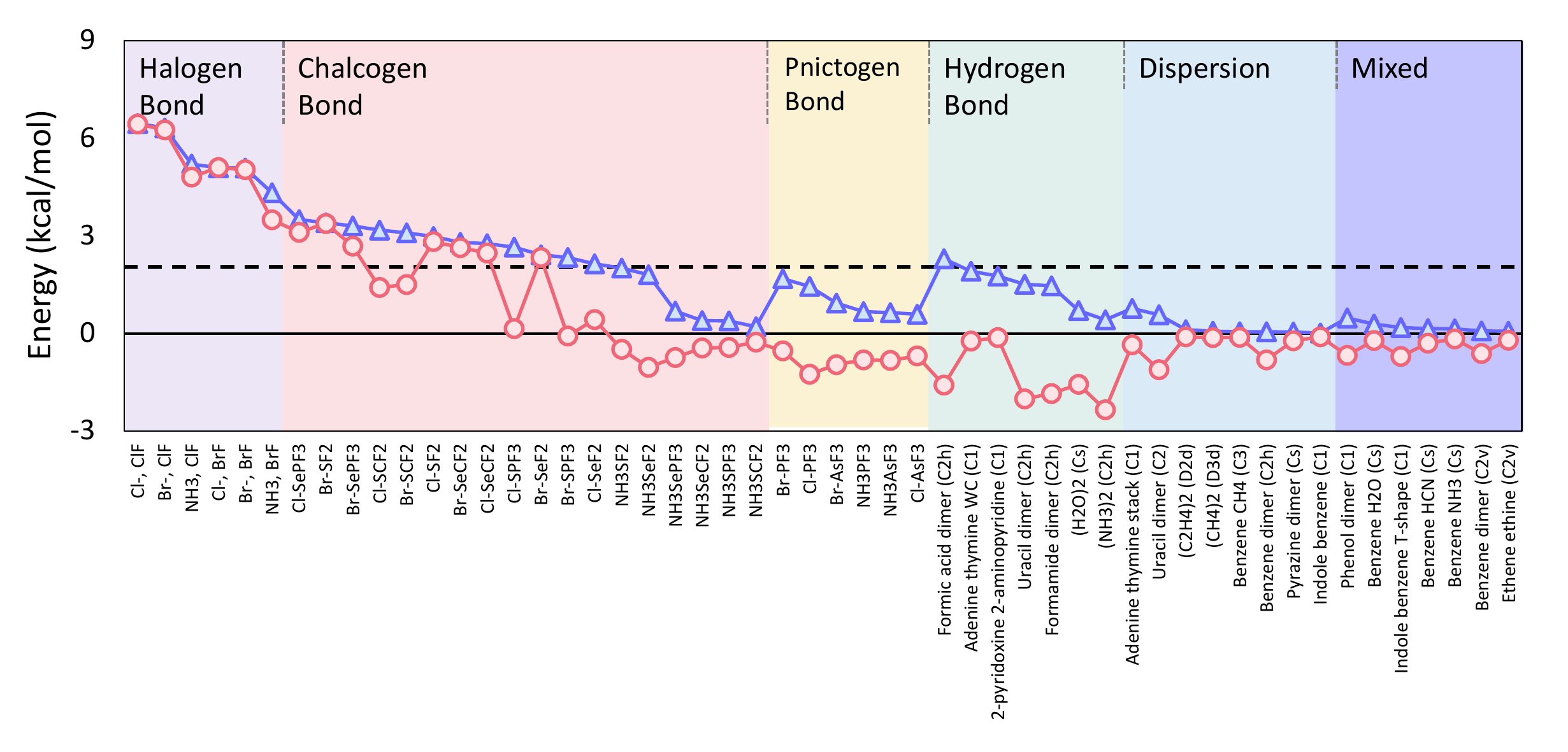}
\caption{
Correlation between density sensitivity ($\tilde S$ of Eq.~\ref{eq:sval}, blue) 
and the difference in absolute reaction energy errors between
self-consistent DFT and HF-DFT (red circles) with PBE
for various non-covalent interactions.
If $S\PBE >$ 2~kcal/mol (dashed horizontal line), the calculation
is density sensitive and DC(HF)-DFT equals to HF-DFT.
%; If red circle > 0, HF density is more accurate than self-consistent density.\cite{SSB18}
DC(HF)-DFT equals to self-consistent DFT for $S\PBE <$2~kcal/mol.  
All calculations use
aug-cc-pVQZ basis set, with
geometries from the B30\cite{BAFE13} and S22\cite{GHBE17,JSCH06} datasets.
}
\label{fgr:s22}
\end{figure*}

\vspace{0.1cm}\noindent{\textbf{\small{\em Problems with indiscriminant use of HF densities}:}}
So, why not use HF densities in all DFT calculations?   The first problem is that self-consistency simplifies tremendously many practical aspects of modern
DFT calculations, such as finding forces, vibration frequencies, polarizabilities and
hyper-polarizabilities, etc.  Anything that can be written as a derivative of the
energy with respect to some parameter becomes much more complicated when the calculation
is not self-consistent.   
The second is
that, in general, if a calculation is not 'abnormal', we have no reason to think
the HF density (or even the exact density)
would yield a more accurate energy than the self-consistent density.
Thus, we may actually reduce accuracy overall if we blindly use HF densities
everywhere.\cite{SVSB22}   
Thirdly, for some difficult systems, 
where the HF calculation is substantially
spin contaminated  
i.e., the HF <$S^2$> is significantly different from the exact value, 
(resulting from an artificial mixing of spin-states\cite{SSH11})
or which are multi-determinantal
in character 
(systems whose physics is poorly described by a single-configuration)\cite{FHPF92}, 
the HF density is likely
to worsen the energy substantially.   Imagine, for example, a database of 100 reaction
energies of some kind.   Suppose, with a given approximate XC, that
5 are abnormal.   And further suppose that using HF densities reduces the abnormal
errors by 5~kcal/mol, on average.  If the HF densities worsen the normal cases by just
0.26~kcal/mol on average, HF-DFT (always using HF densities) worsens the overall results
on the database, and misses the large improvements on the abnormal cases.

\vspace{0.1cm}\noindent{\textbf{\small{\em How to spot when a calculation is density sensitive}:}}
Thus it is crucial to have a procedure or recipe that automatically determines if a
calculation is abnormal.   The original criterion, that the gap is unusually small,
is merely qualitative.\cite{KSB13}
How small is small?   After many variations were tried,
we settled on a simple heuristic, which was called the density sensitivity.\cite{SSB18}  
It is defined as the change in the energy being calculated when going from the 
HF density to the LDA density, where LDA denotes local
density approximation\cite{KS65} (often used in the SVWN form).\cite{VWN80}
\ben
\tilde S = \left| \tilde E[\n\LDA] - \tilde E[\n\HF] \right|.
\label{eq:sval}
\een
This is easily computable in standard molecular codes at DFT cost.
LDA is likely
to suffer more from delocalization than any more modern functional, and so
acts as a canary in a coal mine for density-driven errors.  If $\tilde S$
is significant, we declare the calculation likely to be abnormal
and only then do we use the HF density in place of the self-consistent density.
We found that a cutoff of 2~kcal/mol worked well for most small chemically bonded
molecules, but of course
this value must be adjusted for the circumstances.  It must become
larger for larger molecules,\cite{MH21}
and become smaller for smaller energy differences, such
as in non-covalent interactions\cite{KSSB18}
and intramolecular torsional barriers\cite{NCSB21}.
DC-DFT is the formal name of the analysis 
that leads to these conclusions,\cite{KSB13,KSB14}
and DC(HF)-DFT is the application of HF-DFT only to those cases that are density
sensitive.   Thus, in the 100 reaction energy set, HF-DFT may worsen the overall statistics,
but DC(HF)-DFT will improve them by a small amount, but will produce significant quantitative
improvement on the density-sensitive set.
The theory behind DC-DFT leads to many useful concepts for understanding errors in functionals
and differences between approximate functionals.\cite{VSKS19}

\begin{figure}[htb]
\centering
%\vskip 5cm
\includegraphics[width=1.0\columnwidth]{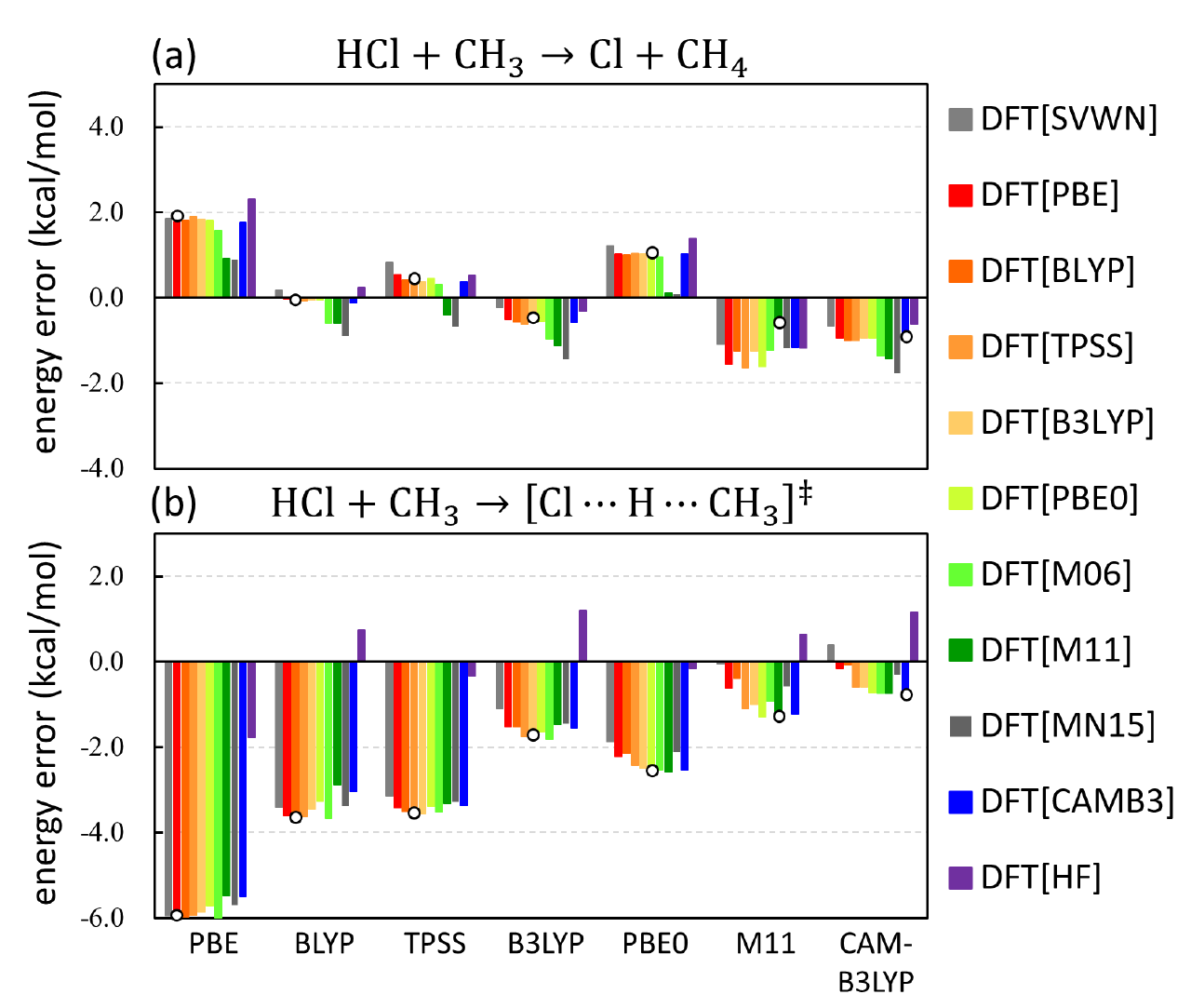}
\caption{
Rainbow plots (several functionals evaluated on each other's densities) for 
(a) $HCl+CH_3 \rightleftarrows Cl+CH_4$ reaction energy
and
(b) its forward barrier height. The
x-axis denotes the XC functional used for calculating energy on different densities (color coded), with
hollow circles marking self-consistency.
All calculations use cc-pVQZ basis, and the
reference is W2-F12 from Ref.~\cite{GHBE17}.
}
\label{fgr:rainbow}
\end{figure}

\vspace{0.1cm}\noindent{\textbf{\small{\em Importance of DC-DFT for non-covalent interactions}:}}
To illustrate the utility of density sensitivity, consider Figure~\ref{fgr:s22}.
Along the $x$-axis, we have listed 52 non-covalently bonded molecules and complexes.
The right-hand-side contains the members of the well-established S22\cite{GHBE17,JSCH06} dataset,
sorted into three categories, depending on
 whether they are hydrogen-bonded, weak dispersion bonds, or mixed.
Within each category, they are arranged in order of PBE density-sensitivity,
with highest on the left.  The PBE sensitivity is the absolute difference between 
the PBE energy on the HF and LDA densities, Eq.~\ref{eq:sval}.   While it increases from right to
left, it only barely reaches 2~kcal/mol for the most sensitive H bonds.  Thus
such weakly bonded compounds are density insensitive, 
and DC(HF)-DFT will not improve
their energetics.   But now look on the left-hand side.  The B30 set contains
unusual weakly-bound molecules in three distinct categories: Pnictogen, chalcogen,
and halogen bonds.\cite{BAFE13}   Overall, their density sensitivities grow from right to
left, and most of the chalcogens and all the halogen cases are density sensitive.
Thus those molecules should have better energetics when HF densities are used.
The red line shows how much the energy error changes when going from the self-consistent
to the HF densities.  It is the difference between the absolute value of the SC error
and the absolute value of the error with HF density.   Where it is positive,
the SC error is larger than the HF error.
Its magnitude tracks the 
%black line 
blue line
very well, showing that
large changes occur where the density sensitivity is largest.  
On the left, the density sensitivity essentially tracks this error difference,
which is positive, and dominated by the self-consistent error.  On the right,
the curves are almost anticorrelated, and the error difference is usually negative,
showing SC yields better energetics than HF densities.
For chalcogens, the mean absolute error of SC-DFT
is 1.9~kcal/mol, while DC(HF)-DFT is 0.5~kcal/mol.\cite{KSSB18}
Note that improvements in energetics for halogen bonds hugely outweigh those due to 
dispersion corrections, as shown in panel (c) of Figure~\ref{fgr:super}.   Thus unwitting inclusion of
such cases into databases for fitting dispersion corrections, without DC(HF)-DFT, worsens
such corrections instead of improving them.\cite{SVSB21}

\vspace{0.1cm}\noindent{\textbf{\small{\em Reducing the spread of DFT results for density-sensitive problems}:}}
Next consider Figure~\ref{fgr:rainbow}, 
which shows many different functionals evaluated on each other's densities
for a simple reaction.   Each collection of bars is the energy of a given functional, using all the
different densities, color coded.  The leftmost bar is gray (LDA density) and
the rightmost is purple (HF density).
The density sensitivity for that functional is the difference between those two.   
In panel (a), the numbers are plotted for the reaction energy.   For any
of the functionals chosen, there is little difference between the gray and purple bars.
This reaction energy is density insensitive.  

But consider panel (b), where the results for forward barrier heights are plotted.
Most functionals give about the same answer, {\em except} when the HF density is used.
Now the differences between gray and purple are huge.  Moreover,
the answer often changes sign (i.e., goes from no barrier in SC-DFT to having a small one in HF-DFT).  Also,
the spread in the different self-consistent answers (shown by open circles) is now far greater than
the variation in the purple bars.   This is a pattern we often see:  If a problem is density sensitive,
often a standard bag of functionals which usually agree with one another show a wide disparity of
results, when evaluated self-consistently.   But on the HF density, their spread is {\em smaller}
than usual.

\vspace{0.1cm}\noindent{\textbf{\small{\em Improved calculations of spin gaps}:}}
The accurate calculation of spin gaps in transition metal complexes is notoriously difficult.
Many methods have very different errors for high- and low-spin states, so calculating their
difference accurately is very difficult.\cite{YN10,K13,FGRT20} It is well-known that many commonly used 
density functionals produce a large spread of answers, much more divergent than they usually
give, especially when mixing HF exchange.\cite{MJMB18}  
Measures of how sensitive the results are on the amount of HF exchange in a DFT functional has been recently used by Kulik and co-workers to improve predictions of properties of transition-metal compounds. \cite{JK17, GK17, VNK22}. 
In extreme cases, even knowing which state is the ground state of a transition-metal complex is difficult.  
Ab initio quantum chemistry also has difficulty in these cases.    Standard CCSD(T) methods
can be converged with very large basis sets, but the usual indicators suggest a strong
multi-reference character, making its reliability questionable.   On the other hand,
multireference methods are difficult to converge with respect to the size of the active
space and the size of ones computer budget.   An alternative approach is to use quantum
Monte Carlo (QMC), a method available for both molecules and solids\cite{BGJA20}, but using totally different
technology to that of {\em ab initio} quantum chemistry.  

\begin{figure*}[htb]
\centering
%\vskip 5cm
\includegraphics[width=1.6\columnwidth]{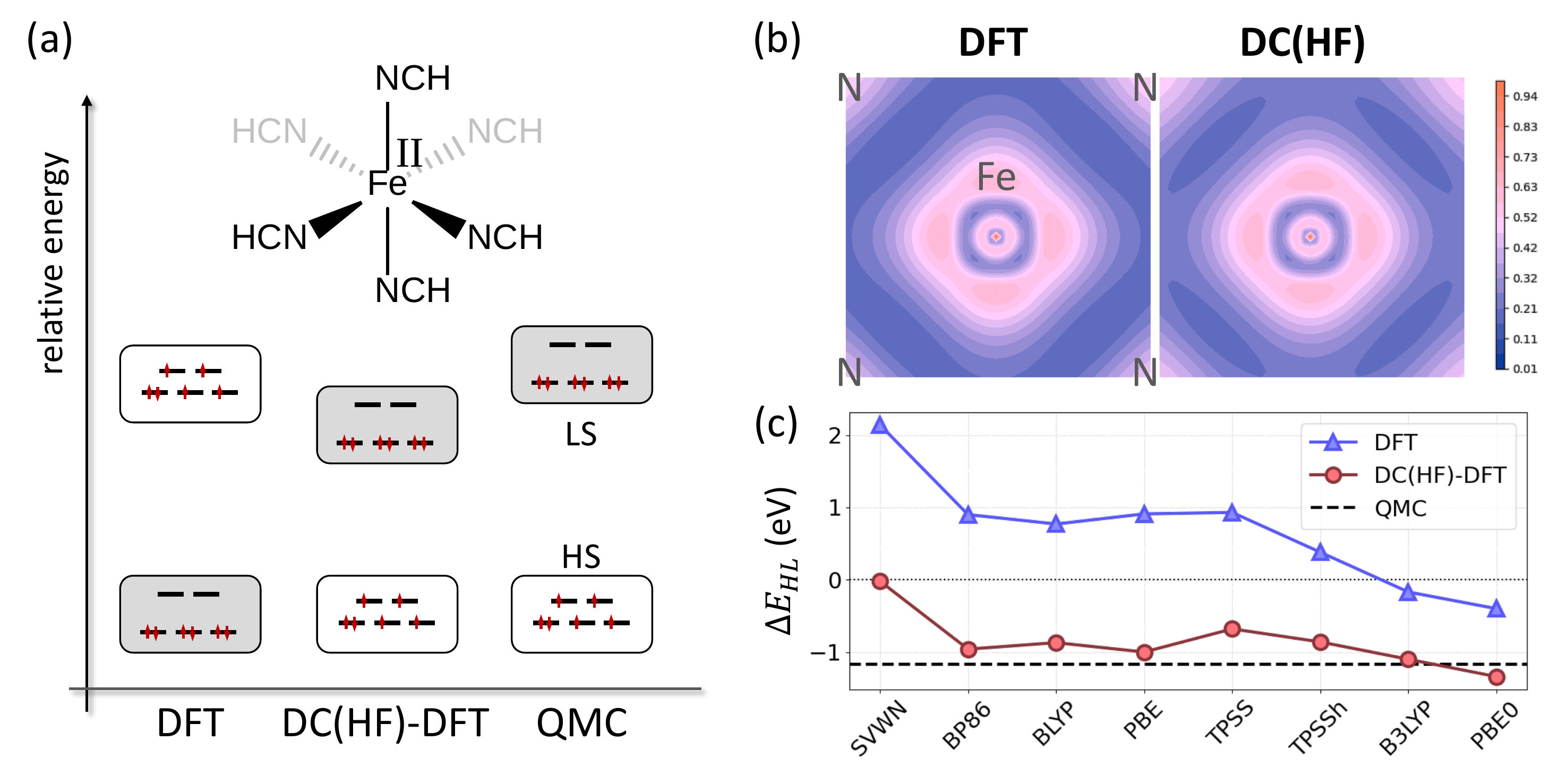}
\caption{
Self-consistent DFT vs.~DC(HF)-DFT results for the spin gap of $Fe[(NCH)_6]^{2+}$ complex.
(a) TPSS and accurate(QMC) ordering of low-spin (LS) and high-spin (HS) states and their relative energies.
(b) Localized-orbital-locator of DFT and HF for the high-spin state. 
(c) Energy difference between high- and low-spin state 
($\Delta E_{\sss HL}=E_{\sss HS}-E_{\sss LS}$)
for various functionals.\cite{SKSB18}  Basis set cc-pVQZ.
}
\label{fgr:spin}
\end{figure*}

Several years ago, a study was performed on 
%6 small molecules with Fe(II) centers and different ligands.\cite{SKSB18}
pseudo octahedral Fe(II) complexes
with various ligands.\cite{SKSB18}
All were wide spin gap cases, of order 1 eV (about 20~kcal/mol) but even so,
different functionals yielded wildly different gaps.   The spread in their results dropped
by about a factor of 2 when HF densities were used instead.  The average results differed
significantly from those of CCSD or CCSD(T), but agreed (within error bars) with some
very expensive, state-of-the-art QMC calculations.   Since then, many authors have tackled
these systems with many variations on many methods, so the jury is still out on whether or
not DFT on HF densities yields accurate spin gaps here.\cite{FGRT20,R19,MVP20}

Figure~\ref{fgr:spin} illustrates some results for the Fe[(NCH)$_6$]$^{2+}$ complex. Panel (a) shows energy
differences between high- and low-spin states.  A metaGGA called
TPSS\cite{TPSS03}, when applied self-consistently, incorrectly yields the
low-spin state as lower than the high-spin state, contradicting
the QMC result.   This is true of many semi-local functionals.  Inclusion of a moderate
fraction of exact exchange may bring the high-spin state slightly lower, but not enough
(see blue curve in (c)).  On the other hand, almost all functionals have the correct
ordering when evaluated on HF densities, and most yield quite accurate spin gaps (red
curve in (c)).  Just as in the rainbow plot of Figure~\ref{fgr:rainbow}, there is a characteristic reduction
in the spread of
predictions when the HF density is used in a density-sensitive system.
Finally, panel (b) shows the localized orbital locator (LOL)\cite{SB00} for both calculations. 
One can see small differences
in the bonding regions using LOL, because it is specifically designed to make such differences
visible, but it is impossible to tell by visual inspection of
densities or their differences which one is better and why.

\begin{figure}[htb]
\centering
%\vskip 5cm
\includegraphics[width=0.8\columnwidth]{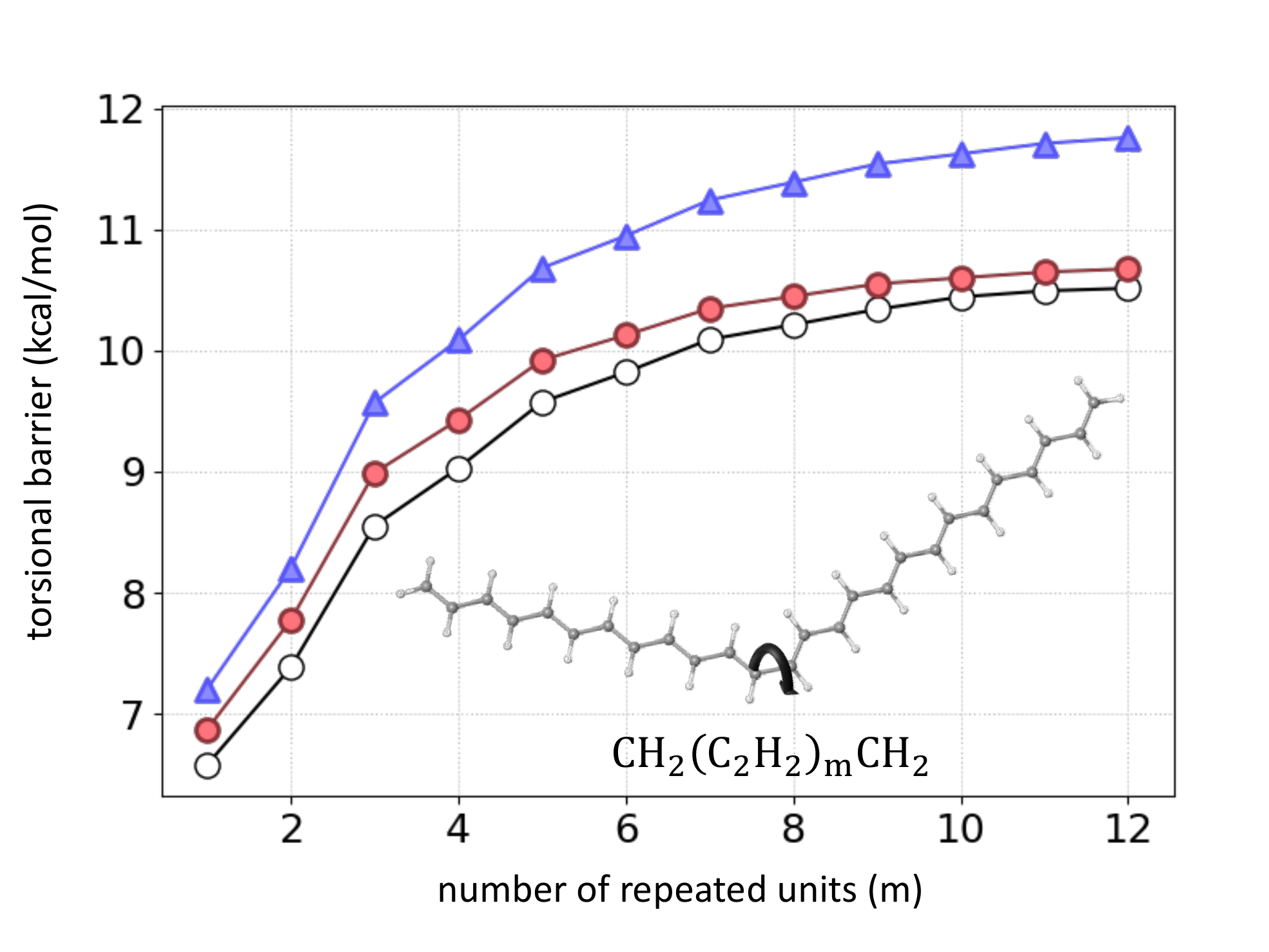}
\caption{
Height of B3LYP torsional barrier of polyacetylene, $CH_2(C_2H_2)_mCH_2$,
for self-consistent DFT (blue triangle) and DC(HF)-DFT (red circle), with RI-MP2-F12 as reference
(open circle aug-cc-pVDZ basis).\cite{NCSB21}
$S>$~2 kJ/mol criterion is used.
}
\label{fgr:poly}
\end{figure}

\vspace{0.1cm}\noindent{\textbf{\small{\em When torsional barrier errors get large}:}}
We return now to the torsional barriers 
in Figure~\ref{fgr:super}(b).  In Ref.~\cite{NCSB21}, %In that work,
the density-sensitivity cutoff was set to 2~kJ/mol instead of 2~kcal/mol, for the
obvious reason that all energetic differences were much smaller than for stronger
chemical interactions.
%processes.   
Nonetheless, the {\em consequences} of errors in self-consistent
DFT torsional barriers can be much larger.   Consider Figure~\ref{fgr:poly}, which shows the
torsional barrier height of conjugated polymer chains at different lengths
%polyacetylene conformers ($CH_2(C_2H_2)_mCH_2$) as a function of $m$, the number of repeated units, 
using the ever-popular B3LYP functional.   The overestimate
of the barrier height grows with 
the chain length,
%$m$,
reaching almost 2~kcal/mol when $m \approx 10$, the number of repeated units.
On the other hand, DC(HF)-B3LYP becomes almost perfect in this limit.

\begin{figure}[htb]
\centering
%\vskip 5cm
\includegraphics[width=0.8\columnwidth]{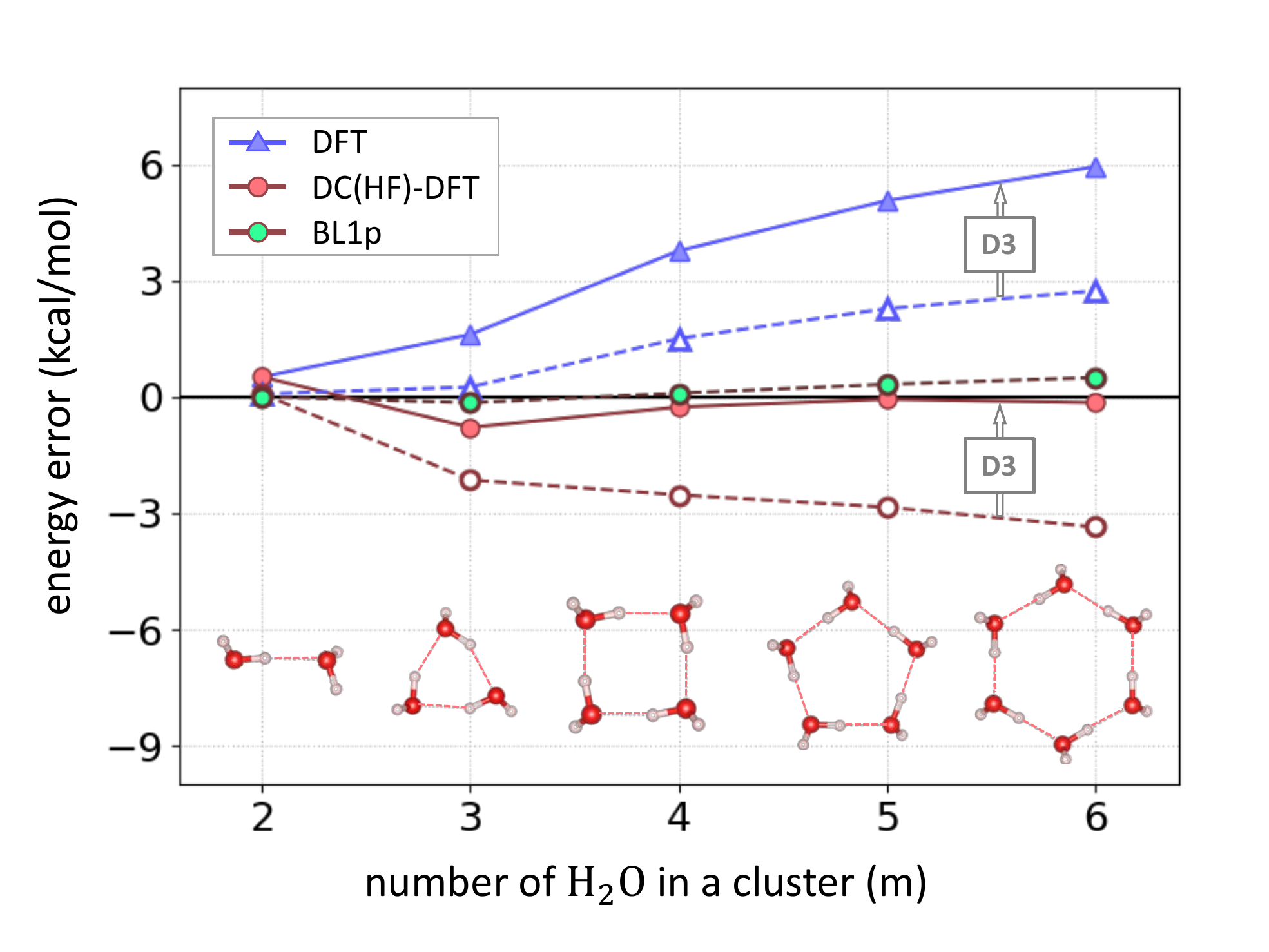}
\caption{
Errors in PBE binding energies
of small water clusters, in self-consistent and DC(HF)-DFT,
with (solid lines) and without (dashed lines) D3 corrections.
DFT calculations use def2-QZVPPD basis set, while the
reference, CCSD(T)-F12/CBS, and geometries are from
the WATER27 dataset.\cite{GHBE17}.
}
\label{fgr:water}
\end{figure}

\setcounter{change}{11}

\vspace{0.1cm}\noindent{\textbf{\small{\em Great success of DC-DFT for water clusters}:}}
To end this tour, we consider binding energies of water clusters.
DFT has been very successful in describing properties of water,\cite{GAM16,LHP21}
and a recent paper has shown that DC-DFT can
achieve near CCSD(T) accuracy for describing a range of water properties,
using the SCAN functional on HF densities.\cite{DLPP21}
In Figure~\ref{fgr:water}, we show errors in PBE binding energies for small
water clusters from self-consistent DFT and DC(HF)-DFT, with and without the D3 correction.
In contrast to the earlier complex shown in Figure~\ref{fgr:super}(c)
where D3 was not affecting the DFT results, here it has a large effect on the DFT errors.
When added to self-consistent DFT, D3 substantially worsens the results, suggesting
an issue inherent to D3.\cite{PBJ21}
But 
in fact D3 greatly reduces the errors when added to
DC(HF)-DFT (in Ref.~\cite{SVSB21}, we discuss in more detail
how large density errors can contaminate and obscure D3 effects). 
BL1p, a double-hybrid which was designed using principles of DC-DFT\cite{SVSB21}
(see next section), is also highly accurate for water complexes.

\section{Theoretical considerations}
\label{theory}

In principle, DC-DFT is a much more general concept than those that appear
in the literature: self-interaction, strong-correlation,
delocalization, straight-line behavior of the energy as a functional of 
non-integer particle number, etc.   It is based on a two-line decomposition of
the error in any DFT calculation.  
Thus it can be applied to every functional approximation ever suggested
and every DFT calculation ever performed, including the first ever
Thomas-Fermi atomic calculations.\cite{T27,F28}   Our focus here has been on KS calculations based
on semi-local approximations, where the HF density typically works to cure significant
density-driven errors, but DC-DFT can be applied much more broadly.   
DC-DFT analysis will usually provide different insights to these traditional analysis
tools and may be less useful.  But more importantly, DC-DFT can unite aspects of these
other characterization tools.

\vspace{0.1cm}\noindent{\textbf{\small{\em Using DC-DFT to quantify errors in densities}:}}
A recent application of DC-DFT involves answering an apparently very simple question:
How do you measure the accuracy of a density?\cite{SSB18}
A popular publication claimed that
some of the most recent empirical density functionals were producing worse densities
than earlier functionals, suggesting that DFT development was `straying from the path'
toward the exact functional.\cite{MBSP17,MBSP17r}
However, closer examination of the methodology used showed
that the results found were due to the choices made by the authors.  Many papers commented
on the original claims\cite{S17,K17c,G17}, some referring to DC-DFT.

With the tools provided by DC-DFT, it is straightforward to address this question from a 
pragmatic viewpoint.   The first and foremost point is that, despite its name, the primary
purpose of (ground-state) DFT is to produce ground-state {\em energies} for different molecular
configurations, not densities.  
Few users ever output or examine the density closely, precisely
because it is not what matters to their results.   Thus the success of DFT in predicting
those energies does not depend on how accurately approximations reproduce the density.   
Nevertheless, when the density or a property computed from it (e.g., electrostatic potential, partial atomic charges, etc.) 
is of interest to a user, 
it is usually better to use HF densities than SC ones provided that a calculation is density sensitive.  
This is illustrated later in Figure~\ref{fgr:nacl},  
where we compare HF and DFT atomic charges as we stretch NaCl.

Of course,
the exact functional reproduces both the exact energy and the exact density but, as we have
seen, a functional which yields usefully accurate energies need not yield accurate densities.
This leads directly to a second important point.   No matter how one might choose to measure
density errors (and there are infinitely many choices, including infinitely many reasonable ones),
there must be some sense of scale.   If density differences are miniscule, why should anybody
care, as they will have essentially no impact on predicted energies?   Thus DC-DFT is the
perfect tool for answering this question, as it measures the accuracy of densities directly
in terms of their impact on the energy.   

\begin{figure}[htb]
\centering
%\vskip 5cm
\includegraphics[width=1.0\columnwidth]{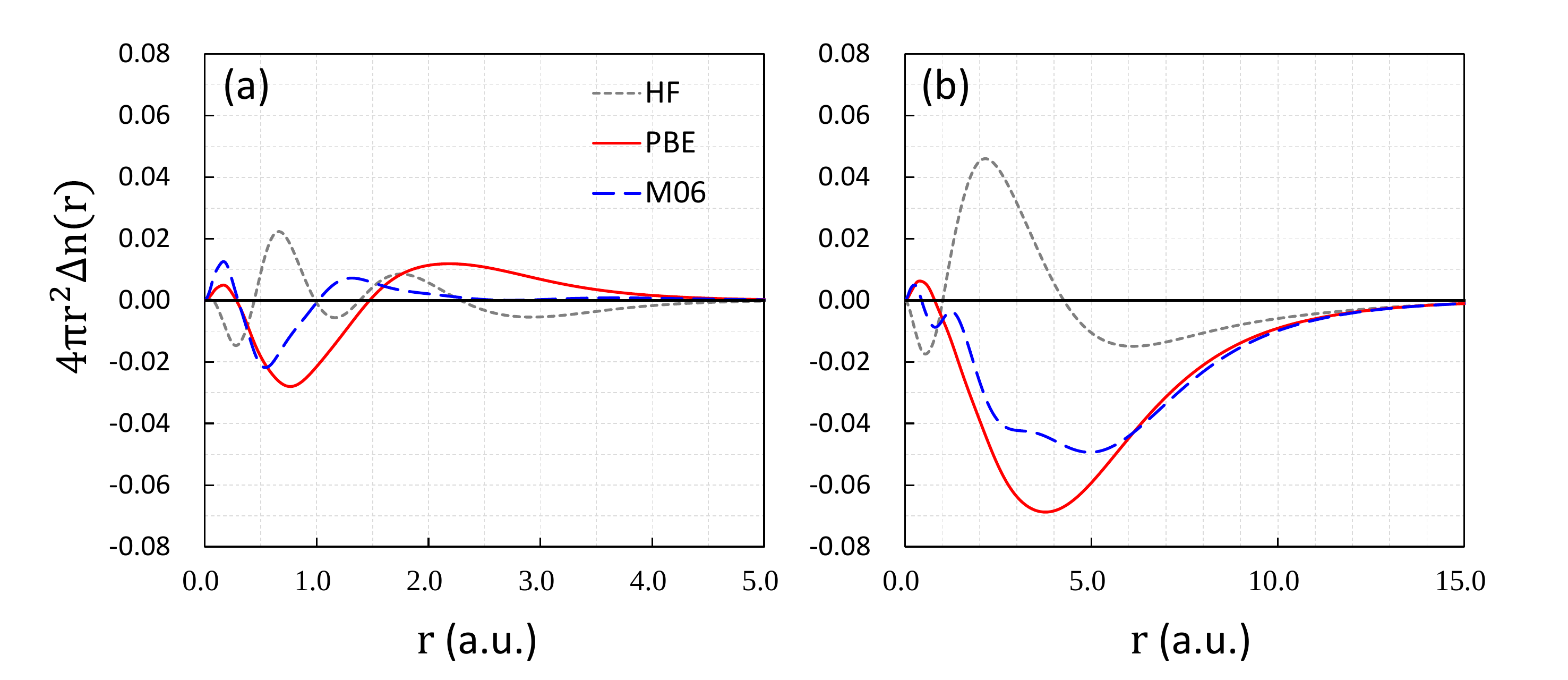}
\caption{
Density errors for a density-insensitive case (a, He atom) and an extremely
density-sensitive case (b, H$^-$), for HF, PBE, and M06.  
An aug-cc-pV5Z basis is used, QMC density references from \cite{UG94},
and convergence method for H$^-$ given in Ref.~\cite{SSB18}, with
0.33 electrons lost by PBE, 0.31 by M06.
}
\label{fgr:density}
\end{figure}

Figure~\ref{fgr:density} shows density errors for two very different two electron
systems.   The first is the He atom which is density-insensitive.  Errors in these
densities have little effect on energies.  Furthermore, given the error profiles,
which approximation has the `best' density?   The ranking depends entirely on one's
choice of measure.   On the right, we have the same errors for H$^-$.   This is
a case of extreme density sensitivity, and was used as the prototype\cite{KSB13} for
understanding density sensitivity.   These are among the `largest' density errors
found in self-consistent DFT calculations, yet they are comparable in magnitude
to those in He.   But the observant will notice that the approximate functionals
(PBE and M06 here) have errors that do not integrate to zero!   How can this be?
In fact, a correct self-consistent calculation\cite{KSB13} has about 0.3 electrons
escape entirely from the system (and a HOMO of exactly zero).   

There are two lessons here.  First, in almost all interesting cases and all those
discussed in this article, the density errors are small and subtle.  We have
never been able to understand energetics from studying these small differences
directly.   The relationship between densities and approximate XC is just far too
complicated.   On the other hand, we never need to do this, as our measures are
all based on calculable energies, which speak for themselves.

The second lesson concerns electron affinities, and is more subtle.   A popular method
for calculating anions with self-consistent semi-local functionals is to
use a basis set similar to that used for the neutral, and find electron affinities by
subtraction.   This method works surprising well, often yielding errors smaller than
those of ionization potentials.  But a sure sign that the anionic calculation is unconverged
is the existence of a positive HOMO.\cite{RT97a,RT97b,JD99}
The basis-set is artificially binding the last electron.
The fully converged calculation is like that of H$^-$ shown, with a fraction of the  last
electron lost to the void.   The beauty of the DC-DFT treatment is that it produces
a well-bound density for the anion without a positive HOMO, and yields accuracies
comparable to the artificial methods in common practice.  In fact, it was studies
of this issue\cite{LB10,LFB10,KSB11}, mostly couched in the language of self-interaction,
that ultimately led to the more general concepts of DC-DFT\cite{KSB13}.

\vspace{0.1cm}\noindent{\textbf{\small{\em Using DC-DFT to avoid altering the fraction of exact exchange}:}}
Becke introduced the idea of a (global) hybrid functional, replacing a fraction
of GGA exchange with HF.\cite{B93,B93HH}
It has since become common to vary the fraction of HF exchange in DFT calculations, both
for molecules and materials.\cite{SDC94, HSE03, YTH04}   
In the molecular case, different functionals are designed
with different amounts.  The original global hybrids had about 20-25~\% exchange, for
reasons that could at least be understood, but more recent (and often more accurate)
functionals might have '2X', or about 50~\% mixing,\cite{ZT06} and many double-hybrid functionals
have even more.\cite{ZXG09,KM13}  But at least the amount is fixed once and for all.  In materials
calculations, it has become increasingly popular to vary the amount of mixing, in order
to position the single-particle levels at some desirable location, such as putting
defect levels correctly in a gap.   Adjusting the amount for each different system
actually leaves the realm of DFT, %density functional theory,
as your functional has picked
up an illegal dependence on the external potential.   Of course, the adjustment may
well be describing good physics, but the road to (formal) hell is paved with good (physical) intuition.
DC-DFT is much less sensitive
to the exchange portion than its self-consistent counterpart
%(compare TPSSh--10~\%, B3LYP--20~\%, and PBE0--25~\%)
providing reliable energies without adjusting the fraction of exact exchange.
In Figure~\ref{fgr:rainbow},  the purple bars differ by about 3 kcal/mol for different functionals,  
but the white dots differ by twice that amount.
This strongly suggest that such adjustments are simply trading density-driven errors for
functional errors, obscuring the underlying physics. Practical DC-DFT never suffers from this
problem because it always uses the same HF density.   Very often, high accuracy is achieved
with the moderate exchange fraction used in popular hybrids such as B3LYP and PBE0.

\vspace{0.1cm}\noindent{\textbf{\small{\em The ease of performing DC-DFT calculations}:}}
As a practical matter, for molecular calculations, it is trivial to evaluate
a density functional on the HF density (and orbitals, if needed).   One
simply converges a HF calculation and then use its solution as the initial
guess in a DFT calculation, while setting the number of iterations to zero.
The computer will evaluate the DFT energy on those orbitals without updating
them.    Scripts for performing this operation are available from the website.\cite{TCCL}

In fact, this is not quite the same as evaluating on the HF density, as the HF
kinetic energy is not quite the same as the KS kinetic energy.\cite{GE95}
However, 
this difference has been found to be much smaller than the improvement typically
provided by using HF densities in cases where density-driven errors are large.\cite{NSSB20}
(See also Figure~\ref{fgr:inversion} below).
In other words, HF densities do far more good than harm for density-sensitive calculations.
Moreover, to the extent practical with finite basis sets, the differences with
using the exact density have been found to also be very small, i.e., use
of HF-DFT yields almost all the benefits that the exact density would confer.

\begin{figure*}[htb]
\centering
%\vskip 5cm
\includegraphics[width=1.95\columnwidth]{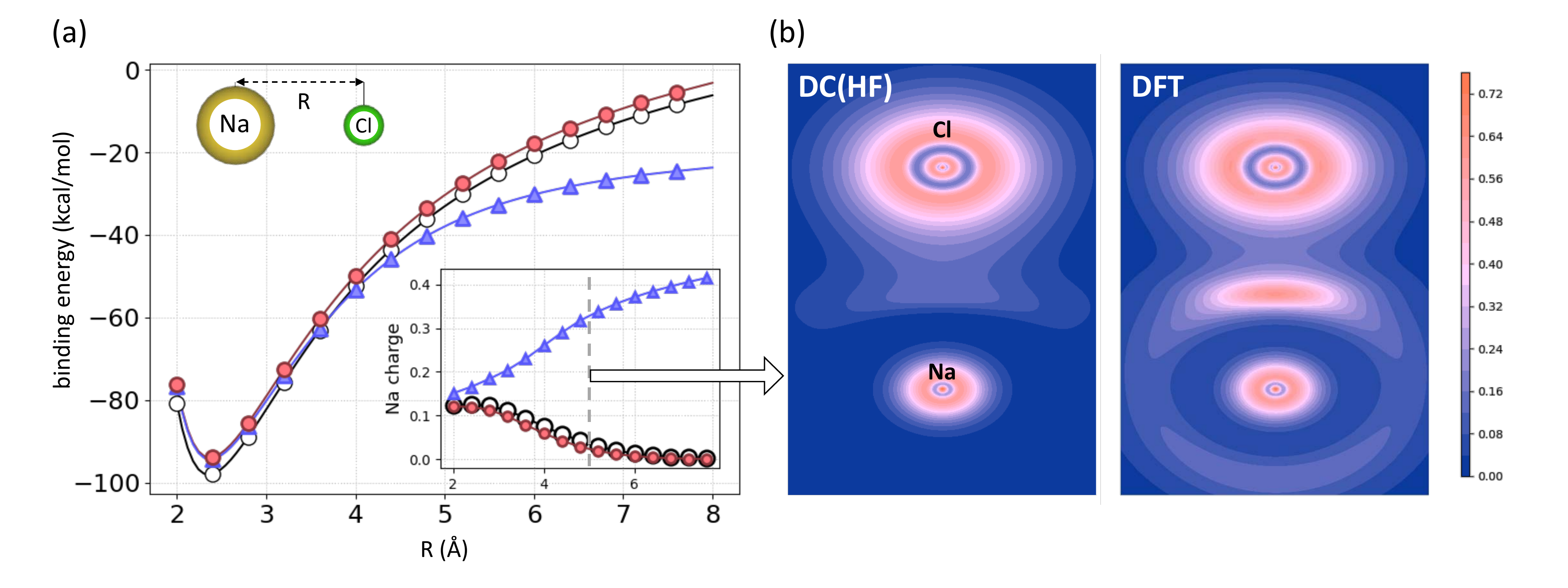}
\caption{
(a) Binding energy of NaCl molecule
from self-consistent DFT (blue triangles) and DC(HF)-DFT (red circles)
compared to reference CCSD(T) calculations(white circles),
using B3LYP in the def2-QZVPPD basis.\cite{KPSS15}
The inset shows the intrinsic atomic orbital charge
of Na atom.
(b) Localized orbital locator of HF (left) and PBE (right)
for NaCl at 5.0~$\AA$.
}
\label{fgr:nacl}
\end{figure*}

\vspace{0.1cm}\noindent{\textbf{\small{\em DC-DFT fixes problems with heteronuclear
stretched bonds}:}}
A major problem with semi-local DFT is failures in binding energy curves.
Typically, as bonds are stretched substantially beyond equilibrium values,
some qualitatively incorrect behavior appears.   For stretched heteronuclear diatomics,
because semi-local functionals are smooth, they allow an incorrect fractional
charge to be transferred, while the exact functional localizes integer
numbers of electrons on each sight. The classic prototype is NaCl(gas),
which dissociated into neutral atoms, unlike NaCl(aq) which dissociates into ions.
In the stretched limit, semi-local DFT tends to unphysically transfer 0.4 electrons
to the Cl ion.  This additional fraction of an electron (and missing
fraction from Na) causes the SC-DFT binding energy curve to be almost 1 eV too
low at large bond distances, as shown in Figure~\ref{fgr:nacl}.  Because HF localizes
charges (more or less correctly, see inset), HF-DFT yields a much more accurate
curve. This correct localization in DC(HF)-DFT can be seen clearly on the right,
where the LOL has been plotted in a plane including the bond axis.

\begin{figure*}[htb]
\centering
%\vskip 5cm
\includegraphics[width=1.95\columnwidth]{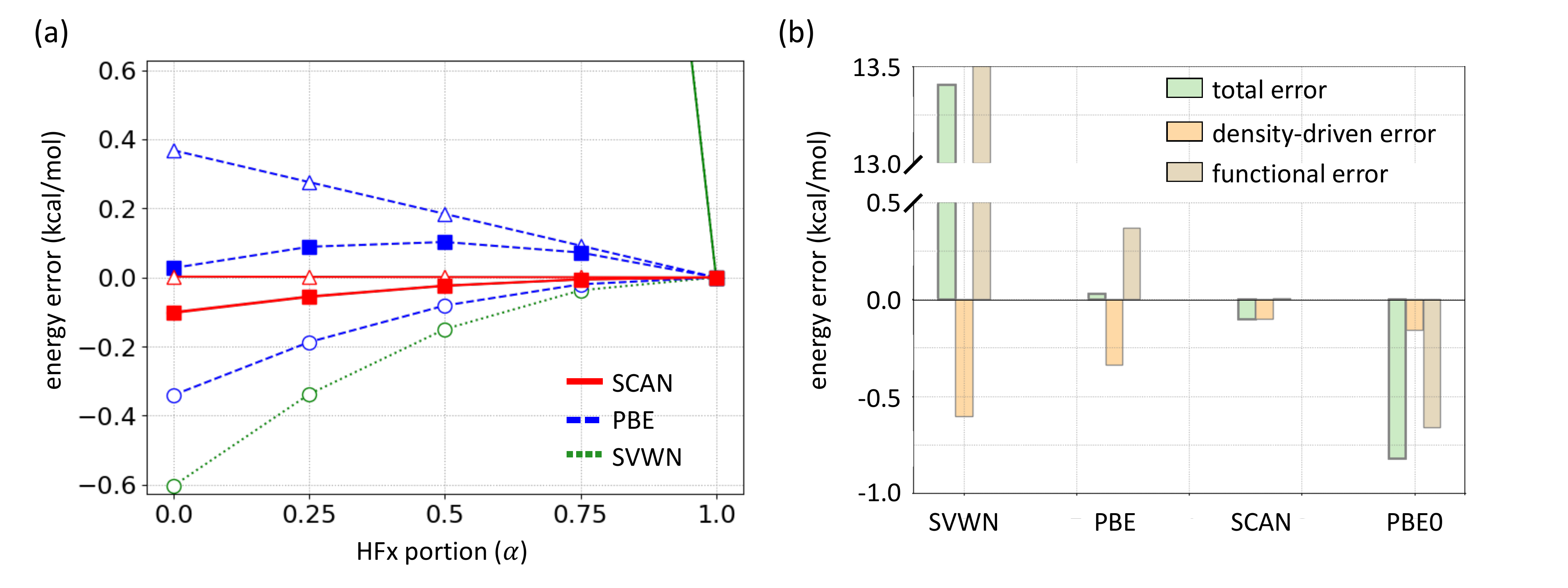}
\caption{
Total errors (squares)
and their components of three approximate functionals on the H atom, as a function
of interpolation to HF (exact here), in the def2-QZVPPD basis. 
At $\alpha=0$, PBE is almost
exact self-consistently, but only due to a cancellation of functional (triangles)
and density-driven errors (circles),
while SCAN is noticeably worse self-consistently, because only its functional error was set to zero.
}
\label{fgr:errdec}
\end{figure*}

\setcounter{change}{16}

\vspace{0.1cm}\noindent{\textbf{\small{\em DC-DFT and functional development}:}}
One way to illustrate the relevance of DC-DFT is to study the evolution of non-empirical functionals
and their global hybrids on the total energy (or ionization energy) of the simplest possible system,
a single H atom.\cite{Ja17,VSKS19,BMDG21}
In Figure~\ref{fgr:errdec}, we consider LDA (SVWN), PBE, and SCAN, and study their behavior under interpolation
toward the exact functional, in this case HF, i.e., $E\DFA\xc+\alpha(E\HF\x-E\DFA\xc)$.
For $\alpha=0$, we have the original functional, but for $\alpha=1$, we have pure HF.
For
$\alpha=0.25$, we have (almost) PBE0 (except correlation has been reduced by 25~\%).  

The functional error of LDA is enormous on this scale, but the density-driven contribution is very small,
illustrating the normalcy of this system and the high typical accuracy of even LDA densities.   But note
the accuracy of the total energy of PBE, i.e., at $\alpha=0$.  That this is accidental can be seen both
by the increased deviation as $\alpha$ grows, but more relevant is that the total error is small because
the functional and density-driven errors are both much larger in magnitude, but mostly cancel.  In fact,
this cancellation is greatest at $\alpha=0$ and is less effective as more of the exact answer is added.
Technically, this makes the H atom density sensitive for PBE, but this is entirely due to the accidental
accuracy at self-consistency.

Finally, we see that SC-SCAN has a larger energy error than SC-PBE for the H atom, but in fact this is 
all density-driven.  The SCAN functional error is zero for the H atom, by construction, but its error
is non-zero when
performed self-consistently.  It can be perfectly possible for an approximate functional to be
designed to be self-interaction error free for exponential densities, and yet produce a finite density-driven
error for the H atom, because it's XC potential will be incorrect.
Ironically, SCAN is less accurate for the H atom than PBE is, despite SCAN using the H atom as an appropriate
norm.

\begin{figure}[htb]
\centering
%\vskip 5cm
\includegraphics[width=0.8\columnwidth]{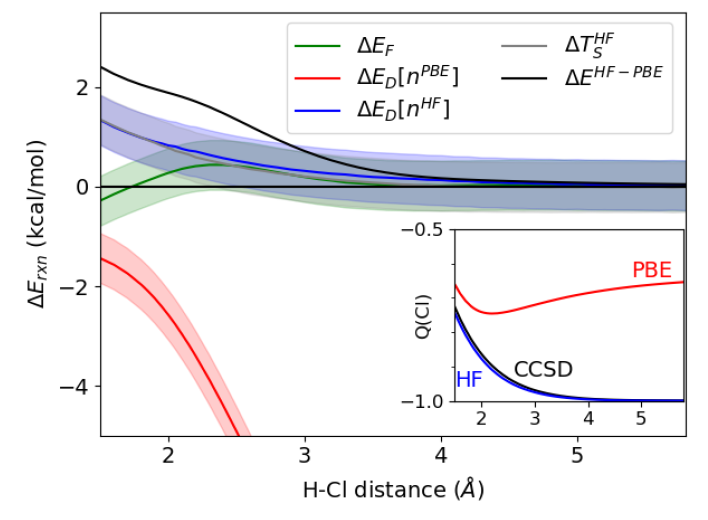}
\caption{
Various energy error components of a PBE OH$\cdot$Cl$^-$ curve.
Inset shows atomic charge on Cl atom.
Reproduced from 
Ref.~\cite{NSSB20}
Copyright 2020 American Chemical Society.
}
\label{fgr:inversion}
\end{figure}

\begin{figure*}[htb]
%\vskip 5cm
\includegraphics[width=1.95\columnwidth]{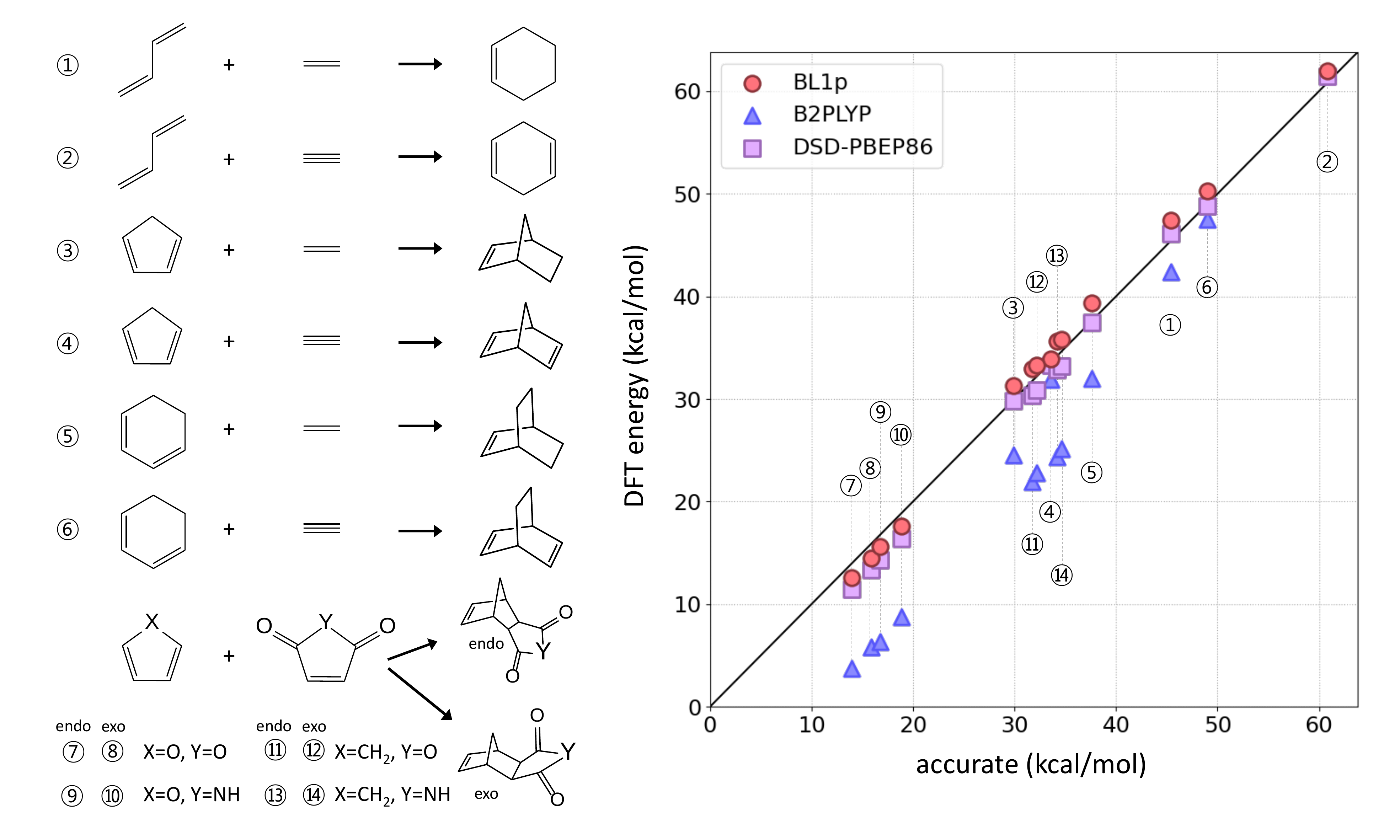}
\caption{
Parity plots for Diels-Alder reaction energies pertaining to the DARC dataset, with
double-hybrids compared against
high-level W1-F12 reference.
BL1p and B2PLYP calculated using def2-QZVPPD basis, while
DSD-PBEP86 is from Ref.~\cite{KM13} and 
reference energies and geometric information on the DARC\cite{JMCY08} dataset are from Ref.~\cite{GHBE17}.
}
\label{fgr:darc}
\end{figure*}

Figure 1(d) showed the PES of self-consistent and DC PBE calculations of OH$\cdot$Cl$^-$.
Figure \ref{fgr:inversion} shows what is happening as OH$\cdot$Cl$^-$ dissociates, breaking down the PBE
errors into their density-driven and functional-driven components, i.e., it is like
Figure~\ref{fgr:errdec}, but looking here at differences.   The shaded regions indicate uncertainties
due to the limitations of KS inversion methods with atomic basis sets, which we use to
reverse-engineer  the 'exact' density and KS orbitals from a correlated wavefunction.\cite{NSSB20}
The density-driven component is sufficiently large as to be off-scale beyond about
2.5 \AA.  Also plotted are the errors of HF-PBE and the small error due to using
the HF kinetic energy instead of the KS kinetic energy.

\vspace{0.1cm}\noindent{\textbf{\small{\em How DC-DFT can improve density-insensitive calculations}:}}
Our last point in the theoretical section concerns functional development.
Double-hybrid functionals, those including fractions of both HF exchange
and MP2 (or other) correlation energy, have been developed and yield
extremely high accuracy, albeit at computational costs greater than traditional
DFT calculations.   In many cases, their densities are so good that one cannot
imagine them suffering from a significant density-driven error.

But, you might be surprised.   The parameters in such functionals are chosen
by minimizing errors on large curated databases, such as the GMTKN55 collection
of 55 databases.\cite{GHBE17}  This process matches the calculated energetic errors relative
the exact energies, not the functional errors.  Moreover, the densities
used in the calculation are typically not quite self-consistent, as it is more involved\cite{SFMB16,FPGG19}
to optimize a functional involving MP2 (which depends on orbital energies).  Thus the
finding of the best parameters has (very small) inefficiencies.   Because these
functionals are so accurate, it only requires a very little density-driven error
to make them suboptimal.

We recently used this insight to demonstrate such issues, creating our own 1-parameter 
double-hybrid, BL1p, but optimized to minimize functional errors rather than total 
energy errors.\cite{SVSB21} 
For standard semilocal density functionals, DC-DFT tells us to use HF densities in density-sensitive cases.   But because of the inclusion of approximate ab initio correlation in double-hybrids, the fraction of exact exchange is typically much higher, 
and it is fine to always use HF density as long as training is done with that density.
The crucial step is to train on the functional error alone, i.e., 
subtracting the density-driven contribution from energy errors.
Thus BL1p fixes the failures of standard double-hybrids for typical density sensitive calculations (e.g., dissociation of NaCl)\cite{SVSB21,SVSB22}, 
but also provides improvements for density insensitive cases as will be illustrated below.
 
In Figure~\ref{fgr:darc}, we show results for the DARC database of Diels-Alder reactions.
These reactions are long known as cases where standard functionals like B3LYP fail badly,
and even double-hybrids.   The figure shows B2PLYP, perhaps the most popular current
double-hybrid, along with the more recent DSD-PBEP86.\cite{KM11}
On the other hand, BL1p uses the exact same ingredients, contains only one empirically
determined parameter 
and uses the HF density. 
Furthermore, BL1p is trained on atomization energies of ony 6 molecules,  while DSD-PBEP86 is trained on many more datapoints. 
The improvement of BL1p over B2PLYP is remarkable,
further reinforcing the need to account for DC-DFT even with the latest, greatest approximations.
We have recently shown that DFT calculations for
DARC reactions are density insensitive making their errors almost entirely
functional-driven.\cite{SVSB21}
Nevertheless, our BL1p still gives improvement for the DARC dataset because
it is designed by the minimization of functional errors, while 
density-driven errors are taken care already by its construction.

\section{Challenges}
\label{chal}

As we have seen, the concept of density-driven errors is becoming widespread in the 
chemical literature and to a lesser extent, in the materials world.\cite{PPSP19,GSVP20,SGMP20}
Moreover, increasing numbers of authors are finding that the selective use of HF densities
does indeed significantly reduce density-driven errors.  In this section, we list some
of the more obvious limitations of the current theory and also where it might be expanded.

\vspace{0.1cm}\noindent{\textbf{\small{\em Stretched H$_2$ and H$_2^+$}:}}
Our first stop is the iconic prototypes of self-interaction error and strong correlation in
chemistry.   These are the binding energy curves of H$_2^+$ and H$_2$, respectively.   The
H$_2^+$ curve is a pure example of unbalanced self-interaction error.  Because it is a one-electron
system, HF densities and energies are exact.  But essentially any semi-local approximation has
an unbalanced error in going from the equilibrium situation to the stretched bond, where
half an electron localizes on each proton.  The self-interaction error changes greatly, leading
to a very unfortunate binding energy curve.   The other example is stretched H$_2$, where the
problem is 1/2 an electron of each spin being localized on each proton.\cite{CMY08,VWMG15} A restricted
KS calculation with a semi-local approximation will dissociate to incorrect fragments with
the wrong energetics (namely, spin-unpolarized H atoms).  

In its current form, DC-DFT has nothing to say about how semi-local DFT can be
improved for these systems, as they do not appear
to be density sensitive.   The errors made by the semi-local approximations on the stretched
bonds are not much different if one uses exact densities or approximate restricted densities.
And evaluation of the approximate functionals on the exact densities still produces the
large errors.   Thus these are functional errors by our current classification scheme.

But did we not say that a success of DC-DFT was to improve the dissociation limit of many
molecules?   Yes we did, but these are heteronuclear molecules whose stretched limit is
not symmetric, and whose HF density is much more accurate in that limit, because of charge
localization.    We suspect that some generalization of DC-DFT ought to be able to 
include both stretched H$_2^+$ and H$_2$ but we have not yet found it.

\vspace{0.1cm}\noindent{\textbf{\small{\em Energy-density consistency}:}}
A second challenge is to restore self-consistency.  
While there are firm theoretical justifications for building DFT 
approximations to be applied to HF densities,\cite{GL94,Detal20,SGVF18,DFDG21}
we have already mentioned the many
practical advantages of using self-consistent densities.
In fact, in principle, restoring self-consistency is always straightforward with
any approximation for the energy.   If we consider $E_N[v]$ as the DC(HF)-DFT energy for
any given problem, characterized by external potential $v(\br)$ and $N$ electrons, then 
the corresponding density is just its functional derivative with respect to $v(\br)$, which 
can always be calculated by making small changes, $v(\br)+\delta v(\br)$ pointwise
in space.   Thus one can imagine performing such a calculation on the DC(HF)-DFT energy.  This
would produce a density that differs from both the HF and the self-consistent (of the original
XC approximation) densities.  
While easy in principle, in practice it may not be, and one could
 use a basis set to represent this density to avoid doing the
calculation pointwise.  It would be very interesting to see in what way such a density
differed from its progenitors, and if it looked more like the exact density.

Of course, a much more satisfactory approach is to construct functionals that yield both
good energetics and good densities when performed self-consistently.\cite{WASJ19}
Perhaps the foremost
approach is that of Yang and co-workers in this direction.\cite{LZSY17} Their
approach is designed to reduce delocalization errors (both functional- and density-driven components)
by explicitly imposing the well-known {\em linearity condition} with respect
to particle number.\cite{PPLB82}  Many failures in DFT are attributed to deviations from this
condition.\cite{KK20}
Cohen and co-workers have recently constructed {\em DeepMind 21}, 
a functional where machine learning has been used to address deviations from linearity condition.\cite{KMTG21}
This and other machine-learning approaches\cite{NAS19,LHPS20} are also very promising when it comes to building functionals that give both good energetics and self-consistent densities. 

\vspace{0.12cm}\vspace{0.1cm}\noindent{\textbf{\small{\em Using DC-DFT to analyse orbital-free DFT errors}:}}
A major use would be to apply it to orbital-free DFT (OF-DFT).  The few
cases we have considered suggest that many orbital-free approximations yield errors that
are dominated by their density-driven component.  When the KS kinetic energy is approximated,
even if very accurately, even small imperfections in the derivative will yield large errors
in the density.\cite{VSKS19}  
Thus DC-DFT allows one to balance improvements 
in the approximate potential (and therefore the density) 
relative to those of the KS kinetic energy functional itself, 
such as whether exact conditions on the potential are relevant to the energies.\cite{RC21}
%While the current form of DC-DFT does not provide a practical form to cure large density-driven errors of OF-DFT, it provides the natural way to quantitatively compare functional- and density-driven errors,
%and thereby guide the development of better orbital-free DFT approximations.} 
This is an area with great potential applications. 
In the light of DC-DFT, another way to view the genius of KS is that it (usually) reduces density-driven errors to negligible amounts.

\vspace{0.1cm}\noindent{\textbf{\small{\em Complications with transition metal chemistry}:}}
There are many more applications of DC-DFT in main group than in transition metal chemistry.
But transition metal applications are slowly catching up.  We illustrated Fe(II) spin
gaps earlier. Also, Martín-Fernández and Harvey applied their {\em normalized}
sensitivity metrics to classify Fe and FeMo clusters by their density sensitivity.\cite{MH21}
In general, more tests are needed to see whether HF densities improve
density-sensitive calculations of transition metal compounds to the extent
they do for main group molecules.  In these cases, it is less clear that the HF density
is sufficiently close to the exact density to guarantee improvement of energetics.

\vspace{0.1cm}\noindent{\textbf{\small{\em Applications to bulk materials and their surfaces}:}}
Another neglected area of application is in materials and surface science.  Almost all our own work has
involved molecular calculations.  In molecular cases, the HF density is easy to calculate, and
is often a good proxy for the exact density in density-sensitive cases.   The need
for DC-DFT analysis may well be even greater in materials calculations than
in molecular calculations.  Do we really know if and when calculations in materials and surface science
suffer from
substantial density-driven errors?  There are a few cases described in the
literature, where the results of semi-local calculations have been analyzed by DC-DFT. 
In some instances (e.g., the adsorption of CO on metallic surfaces), better results
were obtained by the use of presumably more accurate densities.\cite{PPSP19}
In other cases, such as the challenging barrier height for attaching O$_2$ to the Al(111) surface,
semi-local calculations  appear to benefit little from more accurate densities.\cite{GSVP20}
Nevertheless, there are still too few cases of applying DC-DFT in materials
and surface science to draw general conclusions about when and how these fields can benefit from DC-DFT. 
Furthermore,  HF calculations can be formally and computationally problematic in periodic systems.
Janesko overcomes this difficulty by applying DC-DFT without HF exchange.\cite{Ja17}
%This is why a different DC approach may be more useful for these systems and the work of Janesko, where DC-DFT is done differently without HF exchange\cite{Ja17}, could provide a step forward in this direction. } 

\vspace{0.1cm}\noindent{\textbf{\small{\em Forces and geometry optimization with DC-DFT}:}}
From Figure~\ref{fgr:super}(d), we saw that DC-DFT also improves standard DFT forces and
geometries in density-sensitive calculations.  While there are a few codes that
can be used to run geometry optimizations by using DC-DFT\cite{VPB12},
these are not yet widely available.
A more widespread implementation of DC-DFT forces and reactive potentials would facilitate
molecular dynamics based on DC-DFT (to be used in
e.g.,~atmospheric chemistry for odd-electron radical complexes).\cite{KSB14}
Such implementations would make it possible to study
DC-DFT's performance for geometries as well as energies.\cite{VB20}

\vspace{0.1cm}\noindent{\textbf{\small{\em Excited states in DC-DFT}:}}
We are often asked about applying the concepts of DC-DFT to excited states, such as from
the predictions of TDDFT in linear response.\cite{RG84,M16} However,
excited-states do not have their own Hohenberg-Kohn theorem\cite{BG04} and so 
the variational principle upon which so much of DC-DFT is built does not apply here.
On the other hand, there has been a resurgence of interest in ensemble DFT to extract
excited-state energies.\cite{GOK88,YPBU17,GP19}   Ensemble DFT
{\em is} based on a variational principle using the density,  
and so the analysis methods of DC-DFT can be applied.\cite{GP19,G20}

We conclude by simply noting that DC-DFT is based on a simple one-line decomposition
of DFT errors, based on the variational principle.   In the past, many aspects of this
decomposition had been noticed and mused over in understanding DFT results, but DC-DFT
is a formal analysis that puts all these disparate pieces (and disparate sources of
error) together.  The concepts of DC-DFT are appearing more and more frequently in the
chemistry and materials literature, and calculations using DC-DFT are being reported.
As long as researchers continue to use KS-DFT as a standard tool for scientific discovery,
DC-DFT will play an ever-expanding role in analyzing the inevitable errors.
\\

\acknowledgment{
ES and SS are grateful for support from the National Research Foundation of Korea (NRF-2020R1A2C2007468 and NRF-2020R1A4A1017737). 
KB acknowledges funding from NSF (CHEM 1856165).
SV acknowledges funding from the Marie Sklodowska-Curie grant 101033630 (EU’s Horizon 2020 programme).
ES and SS thank Prof. Soo Hyuk Choi for useful comments on the illustrations.
}


\clearpage
\begin{thebibliography}{100}

\bibitem{K99}
Walter Kohn.
\newblock Nobel Lecture: Electronic structure of matter—wave functions and density functionals.
\newblock {\em Reviews of Modern Physics}, 71(5):1253, 1999.

\bibitem{T15}
{TURBOMOLE V7.0 2015}, a development of {University of Karlsruhe} and
  {Forschungszentrum Karlsruhe GmbH}, 1989-2007, {TURBOMOLE GmbH}, since 2007;
  available from {\tt http://www.turbomole.com}.

\bibitem{G16}
M.~J. Frisch, G.~W. Trucks, H.~B. Schlegel, G.~E. Scuseria, M.~A. Robb, J.~R.
  Cheeseman, G.~Scalmani, V.~Barone, G.~A. Petersson, H.~Nakatsuji, X.~Li,
  M.~Caricato, A.~V. Marenich, J.~Bloino, B.~G. Janesko, R.~Gomperts,
  B.~Mennucci, H.~P. Hratchian, J.~V. Ortiz, A.~F. Izmaylov, J.~L. Sonnenberg,
  D.~Williams-Young, F.~Ding, F.~Lipparini, F.~Egidi, J.~Goings, B.~Peng,
  A.~Petrone, T.~Henderson, D.~Ranasinghe, V.~G. Zakrzewski, J.~Gao, N.~Rega,
  G.~Zheng, W.~Liang, M.~Hada, M.~Ehara, K.~Toyota, R.~Fukuda, J.~Hasegawa,
  M.~Ishida, T.~Nakajima, Y.~Honda, O.~Kitao, H.~Nakai, T.~Vreven,
  K.~Throssell, J.~A. Montgomery, {Jr.}, J.~E. Peralta, F.~Ogliaro, M.~J.
  Bearpark, J.~J. Heyd, E.~N. Brothers, K.~N. Kudin, V.~N. Staroverov, T.~A.
  Keith, R.~Kobayashi, J.~Normand, K.~Raghavachari, A.~P. Rendell, J.~C.
  Burant, S.~S. Iyengar, J.~Tomasi, M.~Cossi, J.~M. Millam, M.~Klene, C.~Adamo,
  R.~Cammi, J.~W. Ochterski, R.~L. Martin, K.~Morokuma, O.~Farkas, J.~B.
  Foresman, and D.~J. Fox.
\newblock Gaussian16 {R}evision {B}.01, 2016.
\newblock Gaussian Inc. Wallingford CT.

\bibitem{qchem}
Evgeny Epifanovsky, Andrew~TB Gilbert, Xintian Feng, Joonho Lee, Yuezhi Mao,
  Narbe Mardirossian, Pavel Pokhilko, Alec~F White, Marc~P Coons, Adrian~L
  Dempwolff, et~al.
\newblock Software for the frontiers of quantum chemistry: An overview of
  developments in the q-chem 5 package.
\newblock {\em The Journal of Chemical Physics}, 155(8):084801, 2021.

\bibitem{VASP}
Georg Kresse and J{\"u}rgen Furthm{\"u}ller.
\newblock Efficient iterative schemes for ab initio total-energy calculations
  using a plane-wave basis set.
\newblock {\em Physical Review B}, 54(16):11169, 1996.

\bibitem{PYSCF}
Qiming Sun, Timothy~C. Berkelbach, Nick~S. Blunt, George~H. Booth, Sheng Guo,
  Zhendong Li, Junzi Liu, James~D. McClain, Elvira~R. Sayfutyarova, Sandeep
  Sharma, Sebastian Wouters, and Garnet Kin-Lic Chan.
\newblock Pyscf: the python-based simulations of chemistry framework.
\newblock {\em WIREs Computational Molecular Science}, 8(1):e1340, 2018.

\bibitem{ORCA}
Frank Neese, Frank Wennmohs, Ute Becker, and Christoph Riplinger.
\newblock The orca quantum chemistry program package.
\newblock {\em The Journal of Chemical Physics}, 152(22):224108, 2020.

\bibitem{B12}
Kieron Burke.
\newblock Perspective on density functional theory.
\newblock {\em The Journal of Chemical Physics}, 136(15):150901, 2012.

\bibitem{RCFB08}
Dmitrij Rappoport, Nathan R.~M. Crawford, Filipp Furche, and Kieron Burke.
\newblock {\em Approximate Density Functionals: Which Should I Choose?}
\newblock John Wiley \& Sons, Ltd, 2011.

\bibitem{PGB15}
Aurora Pribram-Jones, David~A. Gross, and Kieron Burke.
\newblock Dft: A theory full of holes?
\newblock {\em Annual Review of Physical Chemistry}, 66(1):283--304, 2015.

\bibitem{MH17}
Narbe Mardirossian and Martin Head-Gordon.
\newblock Thirty years of density functional theory in computational chemistry:
  an overview and extensive assessment of 200 density functionals.
\newblock {\em Molecular Physics}, 115(19):2315--2372, 2017.

\bibitem{GHBE17}
Lars Goerigk, Andreas Hansen, Christoph Bauer, Stephan Ehrlich, Asim Najibi,
  and Stefan Grimme.
\newblock A look at the density functional theory zoo with the advanced gmtkn55
  database for general main group thermochemistry, kinetics and noncovalent
  interactions.
\newblock {\em Physical Chemistry Chemical Physics}, 19(48):32184--32215, 2017.

\bibitem{VT20}
Pragya Verma and Donald~G Truhlar.
\newblock Status and challenges of density functional theory.
\newblock {\em Trends in Chemistry}, 2(4):302--318, 2020.

\bibitem{MS19}
Jan~ML Martin and Golokesh Santra.
\newblock Empirical Double-Hybrid Density Functional Theory: A ‘Third Way’in Between WFT and DFT.
\newblock {\em Israel Journal of Chemistry}, 60:1--19, 2019.

\bibitem{J21}
Benjamin~G. Janesko.
\newblock Replacing hybrid density functional theory: motivation and recent
  advances.
\newblock {\em Chemical Society Reviews}, 50:8470--8495, 2021.

\bibitem{Ja17}
Benjamin~G Janesko.
\newblock Reducing density-driven error without exact exchange.
\newblock {\em Physical Chemistry Chemical Physics}, 19(6):4793--4801, 2017.

\bibitem{NABM19}
Frank Neese, Mihail Atanasov, Giovanni Bistoni, Dimitrios Maganas, and Shengfa
  Ye.
\newblock Chemistry and quantum mechanics in 2019: give us insight and numbers.
\newblock {\em Journal of the American Chemical Society}, 141(7):2814--2824,
  2019.

\bibitem{KS65}
Walter Kohn and Lu~Jeu Sham.
\newblock Self-consistent equations including exchange and correlation effects.
\newblock {\em Physical Review}, 140(4A):A1133, 1965.

\bibitem{HK64}
P.~Hohenberg and W.~Kohn.
\newblock Inhomogeneous electron gas.
\newblock {\em Physical Review}, 136(3B):B864--B871, Nov 1964.

\bibitem{KSB13}
Min-Cheol Kim, Eunji Sim, and Kieron Burke.
\newblock Understanding and reducing errors in density functional calculations.
\newblock {\em Physical Review Letters}, 111(7):073003, 2013.

\bibitem{NCSB21}
Seungsoo Nam, Eunbyol Cho, Eunji Sim, and Kieron Burke.
\newblock Explaining and fixing dft failures for torsional barriers.
\newblock {\em The Journal of Physical Chemistry Letters}, 12(11):2796--2804,
  2021.

\bibitem{KSSB18}
Yeil Kim, Suhwan Song, Eunji Sim, and Kieron Burke.
\newblock Halogen and chalcogen binding dominated by density-driven errors.
\newblock {\em The Journal of Physical Chemistry Letters}, 10(2):295--301,
  2018.

\bibitem{KSB14}
Min-Cheol Kim, Eunji Sim, and Kieron Burke.
\newblock Ions in solution: Density corrected density functional theory
  (dc-dft).
\newblock {\em The Journal of Chemical Physics}, 140(18):18A528, 2014.

\bibitem{LC74}
George~C Lie and Enrico Clementi.
\newblock Study of the electronic structure of molecules. xxi. correlation
  energy corrections as a functional of the hartree-fock density and its
  application to the hydrides of the second row atoms.
\newblock {\em The Journal of Chemical Physics}, 60(4):1275--1287, 1974.

\bibitem{CS75}
Renato Colle and Oriano Salvetti.
\newblock Approximate calculation of the correlation energy for the closed
  shells.
\newblock {\em Theoretica Chimica Acta}, 37(4):329--334, 1975.

\bibitem{GJPF92a}
Peter~MW Gill, Benny~G Johnson, John~A Pople, and Michael~J Frisch.
\newblock An investigation of the performance of a hybrid of hartree-fock and
  density functional theory.
\newblock {\em International Journal of Quantum Chemistry}, 44(S26):319--331,
  1992.

\bibitem{GJPF92b}
Peter~MW Gill, Benny~G Johnson, John~A Pople, and Michael~J Frisch.
\newblock The performance of the Becke—Lee—Yang—Parr (B—LYP) density functional theory with various basis sets.
\newblock {\em Chemical Physics Letters}, 197(4-5):499--505, 1992.

\bibitem{JGP92}
Benny~G Johnson, Peter~MW Gill, and John~A Pople.
\newblock Preliminary results on the performance of a family of density
  functional methods.
\newblock {\em The Journal of Chemical Physics}, 97(10):7846--7848, 1992.

\bibitem{CN93}
Jerzy Cioslowski and Asiri Nanayakkara.
\newblock Electron correlation contributions to one-electron properties from
  functionals of the hartree--fock electron density.
\newblock {\em The Journal of Chemical Physics}, 99(7):5163--5166, 1993.

\bibitem{OB94}
Nevin Oliphant and Rodney~J Bartlett.
\newblock A systematic comparison of molecular properties obtained using
  hartree--fock, a hybrid hartree--fock density-functional-theory, and
  coupled-cluster methods.
\newblock {\em The Journal of Chemical Physics}, 100(9):6550--6561, 1994.

\bibitem{JS08}
Benjamin~G Janesko and Gustavo~E Scuseria.
\newblock Hartree--fock orbitals significantly improve the reaction barrier
  heights predicted by semilocal density functionals.
\newblock {\em The Journal of Chemical Physics}, 128(24):244112, 2008.

\bibitem{CMY08}
Aron~J. Cohen, Paula Mori-S{\'a}nchez, and Weitao Yang.
\newblock Insights into current limitations of density functional theory.
\newblock {\em Science}, 321(5890):792--794, 2008.

\bibitem{LZSY17}
Chen Li, Xiao Zheng, Neil~Qiang Su, and Weitao Yang.
\newblock Localized orbital scaling correction for systematic elimination of
  delocalization error in density functional approximations.
\newblock {\em National Science Review}, 5(2):203--215, 2017.

\bibitem{VPB12}
Prakash Verma, Ajith Perera, and Rodney~J. Bartlett.
\newblock Increasing the applicability of dft i: Non-variational correlation
  corrections from hartree-fock dft for predicting transition states.
\newblock {\em Chemical Physics Letters}, 524:10 -- 15, 2012.

\bibitem{KCK97}
Alfred Karpfen, Cheol~Ho Choi, and Miklos Kertesz.
\newblock Single-bond torsional potentials in conjugated systems: a comparison
  of ab initio and density functional results.
\newblock {\em The Journal of Physical Chemistry A}, 101(40):7426--7433, 1997.

\bibitem{GAEK10}
Stefan Grimme, Jens Antony, Stephan Ehrlich, and Helge Krieg.
\newblock A consistent and accurate ab initio parametrization of density
  functional dispersion correction (dft-d) for the 94 elements h-pu.
\newblock {\em The Journal of Chemical Physics}, 132(15):154104, 2010.

\bibitem{VB20}
Stefan Vuckovic and Kieron Burke.
\newblock Quantifying and understanding errors in molecular geometries.
\newblock {\em The Journal of Physical Chemistry Letters}, 11(22):9957--9964,
  November 2020.

\bibitem{V22}
Stefan Vuckovic.
\newblock Quantification of geometric errors made simple: Application to
  main-group molecular structures.
\newblock {\em The Journal of Physical Chemistry A}, 126(7):1300--1311, 2022.

\bibitem{Pb85}
John~P. Perdew.
\newblock {\em What do the Kohn-Sham orbitals mean? How do atoms dissociate?},
  page 265.
\newblock Plenum, NY, 1985.

\bibitem{PY89}
Robert~G Parr and Weitao Yang.
\newblock {\em Density Functional Theory of Atoms and Molecules}.
\newblock Oxford University Press, 1989.

\bibitem{KPSS15}
Min-Cheol Kim, Hansol Park, Suyeon Son, Eunji Sim, and Kieron Burke.
\newblock Improved dft potential energy surfaces via improved densities.
\newblock {\em The Journal of Physical Chemistry Letters}, 6(19):3802--3807,
  2015.

\bibitem{LZCM15}
Chen Li, Xiao Zheng, Aron~J Cohen, Paula Mori-S{\'a}nchez, and Weitao Yang.
\newblock Local scaling correction for reducing delocalization error in density
  functional approximations.
\newblock {\em Physical Review Letters}, 114(5):053001, 2015.

\bibitem{SSB18}
Eunji Sim, Suhwan Song, and Kieron Burke.
\newblock Quantifying density errors in dft.
\newblock {\em The Journal of Physical Chemistry Letters}, 9(22):6385--6392,
  2018.

\bibitem{VSKS19}
Stefan Vuckovic, Suhwan Song, John Kozlowski, Eunji Sim, and Kieron Burke.
\newblock Density functional analysis: The theory of density-corrected dft.
\newblock {\em Journal of Chemical Theory and Computation}, 15(12):6636--6646,
  2019.

\bibitem{NSSB20}
Seungsoo Nam, Suhwan Song, Eunji Sim, and Kieron Burke.
\newblock Measuring density-driven errors using kohn{\textendash}sham
  inversion.
\newblock {\em Journal of Chemical Theory and Computation}, 16(8):5014--5023,
  July 2020.

\bibitem{SVSB21}
Suhwan Song, Stefan Vuckovic, Eunji Sim, and Kieron Burke.
\newblock Density sensitivity of empirical functionals.
\newblock {\em The Journal of Physical Chemistry Letters}, 12(2):800--807,
  2021.

\bibitem{UG94}
Cyrus~J Umrigar and Xavier Gonze.
\newblock Accurate exchange-correlation potentials and total-energy components
  for the helium isoelectronic series.
\newblock {\em Physical Review A}, 50(5):3827, 1994.

\bibitem{LCB98}
Kin-Chung Lam, Federico~G. Cruz, and Kieron Burke.
\newblock Virial exchange-correlation energy density in hooke{\textquoteright}s
  atom.
\newblock {\em International Journal of Quantum Chemistry}, 69(4):533--540,
  1998.

\bibitem{BCL98}
Kieron Burke, Federico~G. Cruz, and Kin-Chung Lam.
\newblock Unambiguous exchange-correlation energy density.
\newblock {\em J. Chem. Phys.}, 109(19):8161--8167, 1998.

\bibitem{BAFE13}
Antonio Bauza, Ibon Alkorta, Antonio Frontera, and Jose Elguero.
\newblock On the reliability of pure and hybrid dft methods for the evaluation
  of halogen, chalcogen, and pnicogen bonds involving anionic and neutral
  electron donors.
\newblock {\em Journal of Chemical Theory and Computation}, 9(11):5201--5210,
  2013.

\bibitem{JSCH06}
Petr Jure{\v{c}}ka, Ji{\v{r}}{\'\i} {\v{S}}poner, Ji{\v{r}}{\'\i}
  {\v{C}}ern{\`y}, and Pavel Hobza.
\newblock Benchmark database of accurate (mp2 and ccsd (t) complete basis set
  limit) interaction energies of small model complexes, dna base pairs, and
  amino acid pairs.
\newblock {\em Physical Chemistry Chemical Physics}, 8(17):1985--1993, 2006.

\bibitem{SVSB22}
Suhwan Song, Stefan Vuckovic, Eunji Sim, and Kieron Burke.
\newblock Density-corrected dft explained: Questions and answers.
\newblock {\em Journal of Chemical Theory and Computation}, 18(2):817--827,
  2022.

\bibitem{SSH11}
Jason~L. Sonnenberg, H.~Bernhard Schlegel, and Hrant~P. Hratchian.
\newblock {\em Spin Contamination in Inorganic Chemistry Calculations}.
\newblock John Wiley \& Sons, Ltd, 2011.

\bibitem{FHPF92}
James~B Foresman, Martin Head-Gordon, John~A Pople, and Michael~J Frisch.
\newblock Toward a systematic molecular orbital theory for excited states.
\newblock {\em The Journal of Physical Chemistry}, 96(1):135--149, 1992.

\bibitem{VWN80}
Seymour~H Vosko, Leslie Wilk, and Marwan Nusair.
\newblock Accurate spin-dependent electron liquid correlation energies for
  local spin density calculations: a critical analysis.
\newblock {\em Canadian Journal of Physics}, 58(8):1200--1211, 1980.

\bibitem{MH21}
Carlos Martín~Fernández and Jeremy~N. Harvey.
\newblock On the use of normalized metrics for density sensitivity analysis in
  dft.
\newblock {\em The Journal of Physical Chemistry A}, 125(21):4639--4652, 2021.
\newblock PMID: 34018759.

\bibitem{YN10}
Shengfa Ye and Frank Neese.
\newblock Accurate modeling of spin-state energetics in spin-crossover systems
  with modern density functional theory.
\newblock {\em Inorganic Chemistry}, 49(3):772--774, 2010.

\bibitem{K13}
Kasper~P Kepp.
\newblock Consistent descriptions of metal--ligand bonds and spin-crossover in
  inorganic chemistry.
\newblock {\em Coordination Chemistry Reviews}, 257(1):196--209, 2013.

\bibitem{FGRT20}
Benedikt~M Floser, Yang Guo, Christoph Riplinger, Felix Tuczek, and Frank
  Neese.
\newblock Detailed pair natural orbital-based coupled cluster studies of spin
  crossover energetics.
\newblock {\em Journal of Chemical Theory and Computation}, 16(4):2224--2235,
  2020.

\bibitem{MJMB18}
Andrew Mahler, Benjamin~G Janesko, Salvador Moncho, and Edward~N Brothers.
\newblock When hartree-fock exchange admixture lowers dft-predicted barrier
  heights: Natural bond orbital analyses and implications for catalysis.
\newblock {\em The Journal of chemical physics}, 148(24):244106, 2018.

\bibitem{JK17}
Jon~Paul Janet and Heather~J Kulik.
\newblock Predicting electronic structure properties of transition metal
  complexes with neural networks.
\newblock {\em Chemical Science}, 8(7):5137--5152, 2017.

\bibitem{GK17}
Terry~ZH Gani and Heather~J Kulik.
\newblock Unifying exchange sensitivity in transition-metal spin-state ordering
  and catalysis through bond valence metrics.
\newblock {\em Journal of Chemical Theory and Computation}, 13(11):5443--5457,
  2017.

\bibitem{VNK22}
Vyshnavi Vennelakanti, Aditya Nandy, and Heather~J Kulik.
\newblock The effect of hartree-fock exchange on scaling relations and reaction
  energetics for c--h activation catalysts.
\newblock {\em Topics in Catalysis}, 65(1):296--311, 2022.

\bibitem{BGJA20}
Anouar Benali, Kevin Gasperich, Kenneth~D Jordan, Thomas Applencourt, Ye~Luo,
  M~Chandler Bennett, Jaron~T Krogel, Luke Shulenburger, Paul~RC Kent,
  Pierre-Francois Loos, et~al.
\newblock Toward a systematic improvement of the fixed-node approximation in diffusion Monte Carlo for solids—A case study in diamond.
\newblock {\em The Journal of Chemical Physics}, 153(18):184111, 2020.

\bibitem{SKSB18}
Suhwan Song, Min-Cheol Kim, Eunji Sim, Anouar Benali, Olle Heinonen, and Kieron
  Burke.
\newblock Benchmarks and reliable dft results for spin gaps of small ligand fe
  (ii) complexes.
\newblock {\em Journal of Chemical Theory and Computation}, 14(5):2304--2311,
  2018.

\bibitem{R19}
Mariusz Rado{\'n}.
\newblock Benchmarking quantum chemistry methods for spin-state energetics of
  iron complexes against quantitative experimental data.
\newblock {\em Physical Chemistry Chemical Physics}, 21(9):4854--4870, 2019.

\bibitem{MVP20}
Lorenzo~A Mariano, Bess Vlaisavljevich, and Roberta Poloni.
\newblock Biased spin-state energetics of fe (ii) molecular complexes within
  density-functional theory and the linear-response hubbard u correction.
\newblock {\em Journal of Chemical Theory and Computation}, 16(11):6755--6762,
  2020.

\bibitem{TPSS03}
Jianmin Tao, John~P Perdew, Viktor~N Staroverov, and Gustavo~E Scuseria.
\newblock Climbing the density functional ladder: Nonempirical
  meta--generalized gradient approximation designed for molecules and solids.
\newblock {\em Physical Review Letters}, 91(14):146401, 2003.

\bibitem{SB00}
HL~Schmider and AD~Becke.
\newblock Chemical content of the kinetic energy density.
\newblock {\em Journal of Molecular Structure: THEOCHEM}, 527(1-3):51--61,
  2000.

\bibitem{GAM16}
Michael~J Gillan, Dario Alfe, and Angelos Michaelides.
\newblock Perspective: How good is dft for water?
\newblock {\em The Journal of chemical physics}, 144(13):130901, 2016.

\bibitem{LHP21}
Eleftherios Lambros, Jie Hu, and Francesco Paesani.
\newblock Assessing the accuracy of the scan functional for water through a
  many-body analysis of the adiabatic connection formula.
\newblock {\em Journal of Chemical Theory and Computation}, 17(6):3739--3749,
  2021.

\bibitem{DLPP21}
Saswata Dasgupta, Eleftherios Lambros, John~P Perdew, and Francesco Paesani.
\newblock Elevating density functional theory to chemical accuracy for water
  simulations through a density-corrected many-body formalism.
\newblock {\em Nature communications}, 12(1):1--12, 2021.

\bibitem{PBJ21}
Alastair~JA Price, Kyle~R Bryenton, and Erin~R Johnson.
\newblock Requirements for an accurate dispersion-corrected density functional.
\newblock {\em The Journal of Chemical Physics}, 154(23):230902, 2021.

\bibitem{T27}
L.~H. Thomas.
\newblock The calculation of atomic fields.
\newblock {\em Math. Proc. Camb. Phil. Soc.}, 23(05):542--548, 1927.

\bibitem{F28}
E.~Fermi.
\newblock Eine statistische {M}ethode zur {B}estimmung einiger {E}igenschaften
  des {A}toms und ihre {A}nwendung auf die {T}heorie des periodischen {S}ystems
  der {E}lemente (a statistical method for the determination of some atomic
  properties and the application of this method to the theory of the periodic
  system of elements).
\newblock {\em Zeitschrift f\"ur Physik A Hadrons and Nuclei}, 48:73--79, 1928.

\bibitem{MBSP17}
Michael~G Medvedev, Ivan~S Bushmarinov, Jianwei Sun, John~P Perdew, and
  Konstantin~A Lyssenko.
\newblock Density functional theory is straying from the path toward the exact
  functional.
\newblock {\em Science}, 355(6320):49--52, 2017.

\bibitem{MBSP17r}
Michael~G Medvedev, Ivan~S Bushmarinov, Jianwei Sun, John~P Perdew, and
  Konstantin~A Lyssenko.
\newblock Response to Comment on “Density functional theory is straying from the path toward the exact functional”.
\newblock {\em Science}, 356(6337):496--496, 2017.

\bibitem{S17}
Sharon Hammes-Schiffer.
\newblock A conundrum for density functional theory.
\newblock {\em Science}, 355(6320):28--29, 2017.

\bibitem{K17c}
Kasper~P Kepp.
\newblock Comment on `'density functional theory is straying from the path
  toward the exact functional''.
\newblock {\em Science}, 356(6337):496--496, 2017.

\bibitem{G17}
Tim Gould.
\newblock What makes a density functional approximation good? insights from the
  left fukui function.
\newblock {\em Journal of Chemical Theory and Computation}, 13(6):2373--2377,
  2017.

\bibitem{RT97a}
Notker Rosch and SB~Trickey.
\newblock Concerning the applicability of density-functional methods to atomic
  and molecular negative-ions-comment.
\newblock {\em Journal of Chemical Physics}, 106(21):8940--8941, 1997.

\bibitem{RT97b}
Notker R{\"o}sch and SB~Trickey.
\newblock Comment on “Concerning the applicability of density functional methods to atomic and molecular negative ions”[J. Chem. Phys. 105, 862 (1996)].
\newblock {\em The Journal of Chemical Physics}, 106(21):8940--8941, 1997.

\bibitem{JD99}
Andrzej~A Jar{\k{e}}cki and Ernest~R Davidson.
\newblock Density functional theory calculations for f-.
\newblock {\em Chemical Physics Letters}, 300(1-2):44--52, 1999.

\bibitem{LB10}
Donghyung Lee and Kieron Burke.
\newblock Finding electron affinities with approximate density functionals.
\newblock {\em Molecular Physics}, 108(19-20):2687--2701, 2010.

\bibitem{LFB10}
Donghyung Lee, Filipp Furche, and Kieron Burke.
\newblock Accuracy of electron affinities of atoms in approximate density
  functional theory.
\newblock {\em The Journal of Physical Chemistry Letters}, 1(14):2124--2129,
  2010.

\bibitem{KSB11}
Min-Cheol Kim, Eunji Sim, and Kieron Burke.
\newblock Communication: Avoiding unbound anions in density functional
  calculations.
\newblock {\em The Journal of Chemical Physics}, 134(17):171103, 2011.

\bibitem{B93}
Axel~D Becke.
\newblock Density-functional thermochemistry. iii. the role of exact exchange.
\newblock {\em The Journal of Chemical Physics}, 98(7):5648--5652, 1993.

\bibitem{B93HH}
Axel~D Becke.
\newblock A new mixing of hartree--fock and local density-functional theories.
\newblock {\em The Journal of Chemical Physics}, 98(2):1372--1377, 1993.

\bibitem{SDC94}
P.~J. Stephens, F.~J. Devlin, C.~F. Chabalowski, and M.~J. Frisch.
\newblock Ab initio calculation of vibrational absorption and circular
  dichroism spectra using density functional force fields.
\newblock {\em The Journal of Physical Chemistry}, 98(45):11623--11627, 1994.

\bibitem{HSE03}
Jochen Heyd, Gustavo~E Scuseria, and Matthias Ernzerhof.
\newblock Hybrid functionals based on a screened coulomb potential.
\newblock {\em The Journal of Chemical Physics}, 118(18):8207--8215, 2003.

\bibitem{YTH04}
Takeshi Yanai, David~P Tew, and Nicholas~C Handy.
\newblock A new hybrid exchange--correlation functional using the
  coulomb-attenuating method (cam-b3lyp).
\newblock {\em Chemical Physics Letters}, 393(1-3):51--57, 2004.

\bibitem{ZT06}
Yan Zhao and Donald~G Truhlar.
\newblock The m06 suite of density functionals for main group thermochemistry,
  thermochemical kinetics, noncovalent interactions, excited states, and
  transition elements: two new functionals and systematic testing of four
  m06-class functionals and 12 other functionals.
\newblock {\em Theoretical Chemistry Accounts}, 120(1):215--241, 2008.

\bibitem{ZXG09}
Ying Zhang, Xin Xu, and William~A Goddard.
\newblock Doubly hybrid density functional for accurate descriptions of nonbond
  interactions, thermochemistry, and thermochemical kinetics.
\newblock {\em Proceedings of the National Academy of Sciences},
  106(13):4963--4968, 2009.

\bibitem{KM13}
Sebastian Kozuch and Jan~ML Martin.
\newblock Spin-component-scaled double hybrids: an extensive search for the
  best fifth-rung functionals blending dft and perturbation theory.
\newblock {\em Journal of Computational Chemistry}, 34(27):2327--2344, 2013.

\bibitem{TCCL}
Yonsei~University Theoretical and Computational~Chemistry Laboratory.
\newblock Density corrected-density functional theory.
\newblock \url{http://tccl.yonsei.ac.kr/mediawiki/index.php/DC-DFT}.
\newblock visited on 2022-03-14.

\bibitem{GE95}
Andreas G{\"o}rling and Matthias Ernzerhof.
\newblock Energy differences between kohn-sham and hartree-fock wave functions
  yielding the same electron density.
\newblock {\em Physical Review A}, 51(6):4501, 1995.

\bibitem{BMDG21}
Hugh G.~A. Burton, Clotilde Marut, Timothy~J. Daas, Paola Gori-Giorgi, and
  Pierre-Francois Loos.
\newblock Variations of the Hartree–Fock fractional-spin error for one electron.
\newblock {\em The Journal of Chemical Physics}, 155(5):054107, 2021.

\bibitem{JMCY08}
Erin~R Johnson, Paula Mori-S{\'a}nchez, Aron~J Cohen, and Weitao Yang.
\newblock Delocalization errors in density functionals and implications for
  main-group thermochemistry.
\newblock {\em The Journal of Chemical Physics}, 129(20):204112, 2008.

\bibitem{SFMB16}
Szymon {\'S}miga, Odile Franck, Bastien Mussard, Adam Buksztel, Ireneusz
  Grabowski, Eleonora Luppi, and Julien Toulouse.
\newblock Self-consistent double-hybrid density-functional theory using the
  optimized-effective-potential method.
\newblock {\em The Journal of Chemical Physics}, 145(14):144102, 2016.

\bibitem{FPGG19}
Eduardo Fabiano, Szymon {\'{S}}miga, Sara Giarrusso, Timothy~J Daas, Fabio
  Della~Sala, Ireneusz Grabowski, and Paola Gori-Giorgi.
\newblock Investigation of the exchange-correlation potentials of functionals
  based on the adiabatic connection interpolation.
\newblock {\em Journal of Chemical Theory and Computation}, 15(2):1006--1015,
  2019.

\bibitem{KM11}
Sebastian Kozuch and Jan~ML Martin.
\newblock Dsd-pbep86: in search of the best double-hybrid dft with
  spin-component scaled mp2 and dispersion corrections.
\newblock {\em Physical Chemistry Chemical Physics}, 13(45):20104--20107, 2011.

\bibitem{PPSP19}
Abhirup Patra, Haowei Peng, Jianwei Sun, and John~P Perdew.
\newblock Rethinking co adsorption on transition-metal surfaces: Effect of
  density-driven self-interaction errors.
\newblock {\em Physical Review B}, 100(3):035442, 2019.

\bibitem{GSVP20}
Nick Gerrits, Egidius W.~F. Smeets, Stefan Vuckovic, Andrew~D. Powell,
  Katharina Doblhoff-Dier, and Geert-Jan Kroes.
\newblock Density functional theory for molecule{\textendash}metal surface
  reactions: When does the generalized gradient approximation get it right, and
  what to do if it does not.
\newblock {\em The Journal of Physical Chemistry Letters}, 11(24):10552--10560,
  December 2020.

\bibitem{SGMP20}
Jonathan~M. Skelton, David S.~D. Gunn, Sebastian Metz, and Stephen~C. Parker.
\newblock Accuracy of hybrid functionals with non-self-consistent
  kohn{\textendash}sham orbitals for predicting the properties of
  semiconductors.
\newblock {\em Journal of Chemical Theory and Computation}, 16(6):3543--3557,
  May 2020.

\bibitem{VWMG15}
Stefan Vuckovic, Lucas~O Wagner, Andr{\'e} Mirtschink, and Paola Gori-Giorgi.
\newblock Hydrogen molecule dissociation curve with functionals based on the
  strictly correlated regime.
\newblock {\em Journal of Chemical Theory and Computation}, 11(7):3153--3162,
  2015.

\bibitem{GL94}
A.~G{\"o}rling and M.~Levy.
\newblock Exact kohn-sham scheme based on perturbation theory.
\newblock {\em Physical Review A}, 50:196, 1994.

\bibitem{Detal20}
Timothy~J Daas, Juri Grossi, Stefan Vuckovic, Ziad~H Musslimani, Derk~P Kooi,
  Michael Seidl, Klaas~JH Giesbertz, and Paola Gori-Giorgi.
\newblock Large coupling-strength expansion of the m{\o}ller--plesset adiabatic
  connection: From paradigmatic cases to variational expressions for the
  leading terms.
\newblock {\em The Journal of Chemical Physics}, 153(21):214112, 2020.

\bibitem{SGVF18}
Michael Seidl, Sara Giarrusso, Stefan Vuckovic, Eduardo Fabiano, and Paola
  Gori-Giorgi.
\newblock Communication: Strong-interaction limit of an adiabatic connection in
  hartree-fock theory.
\newblock {\em The Journal of chemical physics}, 149(24):241101, 2018.

\bibitem{DFDG21}
Timothy~J Daas, Eduardo Fabiano, Fabio Della~Sala, Paola Gori-Giorgi, and
  Stefan Vuckovic.
\newblock Noncovalent interactions from models for the m{\o}ller--plesset
  adiabatic connection.
\newblock {\em The journal of physical chemistry letters}, 12:4867--4875, 2021.

\bibitem{WASJ19}
Kushantha~PK Withanage, Sharmin Akter, Chandra Shahi, Rajendra~P Joshi, Carlos
  Diaz, Yoh Yamamoto, Rajendra Zope, Tunna Baruah, John~P Perdew, Juan~E
  Peralta, et~al.
\newblock Self-interaction-free electric dipole polarizabilities for atoms and
  their ions using the fermi-l{\"o}wdin self-interaction correction.
\newblock {\em Physical Review A}, 100(1):012505, 2019.

\bibitem{PPLB82}
John~P. Perdew, Robert~G. Parr, Mel Levy, and Jose~L. Balduz.
\newblock Density-functional theory for fractional particle number: Derivative
  discontinuities of the energy.
\newblock {\em Physical Review Letters}, 49:1691--1694, Dec 1982.

\bibitem{KK20}
Leeor Kronik and Stephan K{\"u}mmel.
\newblock Piecewise linearity, freedom from self-interaction, and a coulomb
  asymptotic potential: three related yet inequivalent properties of the exact
  density functional.
\newblock {\em Physical Chemistry Chemical Physics}, 22(29):16467--16481, 2020.

\bibitem{KMTG21}
James Kirkpatrick, Brendan McMorrow, David~HP Turban, Alexander~L Gaunt,
  James~S Spencer, Alexander~GDG Matthews, Annette Obika, Louis Thiry, Meire
  Fortunato, David Pfau, et~al.
\newblock Pushing the frontiers of density functionals by solving the
  fractional electron problem.
\newblock {\em Science}, 374(6573):1385--1389, 2021.

\bibitem{NAS19}
Ryo Nagai, Ryosuke Akashi, and Osamu Sugino.
\newblock Completing density functional theory by machine learning hidden
  messages from molecules.
\newblock {\em npj Computational Materials}, 6(1):1--8, 2020.

\bibitem{LHPS20}
Li~Li, Stephan Hoyer, Ryan Pederson, Ruoxi Sun, Ekin~D. Cubuk, Patrick Riley,
  and Kieron Burke.
\newblock Kohn-sham equations as regularizer: Building prior knowledge into
  machine-learned physics.
\newblock {\em Physical Review Letters}, 126:036401, Jan 2021.

\bibitem{RC21}
Jeremy~J Redd and Antonio~C Cancio.
\newblock Analysis of atomic pauli potentials and their large-z limit.
\newblock {\em The Journal of Chemical Physics}, 155(13):134112, 2021.

\bibitem{RG84}
Erich Runge and E.~K.~U. Gross.
\newblock Density-functional theory for time-dependent systems.
\newblock {\em Phys. Rev. Lett.}, 52(12):997, Mar 1984.

\bibitem{M16}
Neepa~T Maitra.
\newblock Perspective: Fundamental aspects of time-dependent density functional
  theory.
\newblock {\em The Journal of Chemical Physics}, 144(22):220901, 2016.

\bibitem{BG04}
P.~Bokes and R.~W. Godby.
\newblock Conductance and polarization in quantum junctions.
\newblock {\em Phys. Rev. B}, 69:245420, 2004.

\bibitem{GOK88}
Eberhardt~KU Gross, Luiz~N Oliveira, and Walter Kohn.
\newblock Density-functional theory for ensembles of fractionally occupied
  states. i. basic formalism.
\newblock {\em Physical Review A}, 37(8):2809, 1988.

\bibitem{YPBU17}
Zeng-hui Yang, Aurora Pribram-Jones, Kieron Burke, and Carsten~A. Ullrich.
\newblock Direct extraction of excitation energies from ensemble
  density-functional theory.
\newblock {\em Phys. Rev. Lett.}, 119:033003, Jul 2017.

\bibitem{GP19}
Tim Gould and Stefano Pittalis.
\newblock Density-driven correlations in many-electron ensembles: Theory and
  application for excited states.
\newblock {\em Physical Review Letters}, 123(1):016401, 2019.

\bibitem{G20}
Tim Gould.
\newblock Approximately self-consistent ensemble density functional theory:
  toward inclusion of all correlations.
\newblock {\em The Journal of Physical Chemistry Letters}, 11(22):9907--9912,
  2020.

\end{thebibliography}
\end{document}